\documentclass[aps,prx,reprint,superscriptaddress,floatfix]{revtex4-2}
\pdfoutput=1

\usepackage{silence}
\usepackage[english]{babel}

\usepackage[T1]{fontenc}
\usepackage{stix2}
\usepackage{upgreek}
\newcommand{\upDelta}{\Updelta}

\usepackage[version=4,arrows=pgf]{mhchem}
\usepackage{chemmacros}

\usepackage{siunitx}

\usepackage[stretch=15,shrink=15,step=1]{microtype}

\usepackage{xcolor}

\usepackage{nicefrac}

\usepackage{checkend}

\usepackage{graphicx}

\usepackage{float, flafter}

\usepackage{booktabs}

\usepackage[nolist,nohyperlinks]{acronym}

\usepackage{mleftright}

\usepackage{xfp}

\usepackage{fancyhdr}

\usepackage{bibunits}
\usepackage{natbib}

\usepackage[bookmarks, pdftitle={Simulations of DNA-origami self-assembly reveal design-dependent nucleation barriers}, pdfauthor={Alexander Cumberworth, Daan Frenkel, Aleks Reinhardt}, pdffitwindow=false, pdfstartview=FitH, pdfdisplaydoctitle, colorlinks, plainpages=false, pdfpagelabels, hypertexnames, citecolor={blue}, linkcolor={teal}, urlcolor={violet}, pdflang={en}, hyperfootnotes=false, breaklinks]{hyperref}

\usepackage[capitalise,noabbrev]{cleveref}

\definecolor{textblack}{gray}{0.125}

\allowdisplaybreaks

\fancypagestyle{main}{
    \fancyhf{}

    \fancyfoot[C]{\thepage}
    \pagenumbering{arabic}
}

\input{tex/macros_arxiv}

\graphicspath{{figs-final2/}}

\begin{document}
    \title{\textcolor{textblack}{Simulations of DNA-origami self-assembly reveal design-dependent nucleation barriers}}
    \author{\textcolor{textblack}{Alexander Cumberworth}}
    \email{alex@cumberworth.org}
    \affiliation{\textcolor{textblack}{AMOLF, Science Park 104, 1098 XG Amsterdam, Netherlands}}
    \author{\textcolor{textblack}{Daan Frenkel}}
    \author{\textcolor{textblack}{Aleks Reinhardt}}
    \affiliation{\textcolor{textblack}{Yusuf Hamied Department of Chemistry, University of Cambridge, Lensfield Road, Cambridge, CB2 1EW, United Kingdom}}
    \defaultcolor{textblack}


\begin{acronym}
    \acro{bp}{base pair}
    \acro{CB}{configurational bias}
    \acro{CT}{conserved topology}
    \acro{CTRG}{conserved-topology recoil-growth}
    \acro{LFE}{Landau free energy}
    \acroplural{LFE}[LFEs]{Landau free energies}
    \acro{MBAR}{multistate Bennett acceptance ratio}
    \acro{MC}{Monte Carlo}
    \acro{NN}{nearest-neighbour}
    \acro{nt}{nucleotide}
    \acro{REMC}{replica-exchange Monte Carlo}
    \acro{RG}{recoil-growth}
    \acro{SI}{Supporting Information}
    \acro{US}{umbrella sampling}
    \acro{MWUS}{multi-window umbrella sampling}
    \acro{REMWUS}{replica-exchange multi-window umbrella sampling}
    \acro{WHAM}{weighted-histogram analysis method}
\end{acronym}

    \begin{abstract}
        Nucleation is the rate-determining step in the kinetics of many self-assembly processes.
However, the importance of nucleation in the kinetics of DNA-origami self-assembly, which involves both the binding of staple strands and the folding of the scaffold strand, is unclear.
Here, using Monte Carlo simulations of a lattice model of DNA origami, we find that some, but not all, designs can have a nucleation barrier and that this barrier disappears at lower temperatures, rationalizing the success of isothermal assembly.
We show that the height of the nucleation barrier depends primarily on the coaxial stacking of staples that are adjacent on the same helix, a parameter that can be modified with staple design.
Creating a nucleation barrier to DNA-origami assembly could be useful in optimizing assembly times and yields, while eliminating the barrier may allow for fast molecular sensors that can assemble/disassemble without hysteresis in response to changes in the environment.

    \end{abstract}

    \maketitle
    \defaultcolor{textblack}
    \pagestyle{main}
    The design and production of DNA-origami structures has grown into a mature field~\cite{seeman2015}.
In these structures, a long DNA `scaffold' strand is folded into a target structure by hybridizing with a number of designed shorter `staple' strands that connect chosen binding domains on the scaffold strand.
However, while there is much practical knowledge on how to optimize the assembly of DNA origamis~\cite{rossi-gendron2022,halley2019,hong2017}, an understanding of the underlying physical mechanisms, such as the nature of any free-energy barriers to assembly and their dependence on assembly conditions, is lacking.

There is some experimental evidence that nucleation may be less important for origami self-assembly than for other assembly processes, such as crystallization.
For instance, although DNA-origami assembly is often performed by slowly decreasing the temperature of a mixture with an excess of staple strands over several hours or even days~\cite{wagenbauer2017}, it is also possible to assemble such structures isothermally following a high-temperature denaturing step~\cite{jungmann2008,sobczak2012,song2013,zhang2013b,kopielski2015,song2017,halley2019,schneider2019,rossi-gendron2022}.
Moreover, isothermal assembly has been shown to be faster for a range of designs, with the optimal temperature for this process depending on both the design of the target structure and the conditions~\cite{sobczak2012,halley2019,rossi-gendron2022}.
On the other hand, many studies on DNA origami have found hysteresis between melting and annealing as the temperature is varied~\cite{kosinski2019,wah2016,shapiro2015,dunn2015,dannenberg2015,arbona2013,wei2013,arbona2012,arbona2012b,sobczak2012}, which suggests the presence of significant free-energy barriers.
It has been suggested that the melting--annealing hysteresis could be attributed to a nucleation barrier to staple binding~\cite{schneider2019,ke2012b,sobczak2012}, but no numerical evidence has been given to show that such a barrier exists.

In contrast to DNA-origami assembly, nucleation has been shown to be important in the self-assembly of `DNA-brick' structures~\cite{ke2012,wei2012}, which consist of a large number of short unique strands that assemble in the absence of a scaffold strand.
The nucleation barrier for DNA-brick self-assembly plays an important role in allowing error-free assembly of these many-component systems~\cite{reinhardt2014}.
This barrier has been studied in some depth, as control of the nucleation barrier enables the design of DNA-brick structures that have favourable assembly kinetics~\cite{jacobs2015,reinhardt2016,wayment-steele2017,sajfutdinow2018,fonseca2018,zhang2020}.
By contrast, although DNA-origami self-assembly has been successfully modelled~\cite{dannenberg2015,dunn2015,snodin2016} and subsequently validated~\cite{marras2016}, most existing simulation methods are too computationally expensive to allow for a systematic study of possible nucleation barriers.

\begin{figure*}[!ht]
    \centering
    \includegraphics{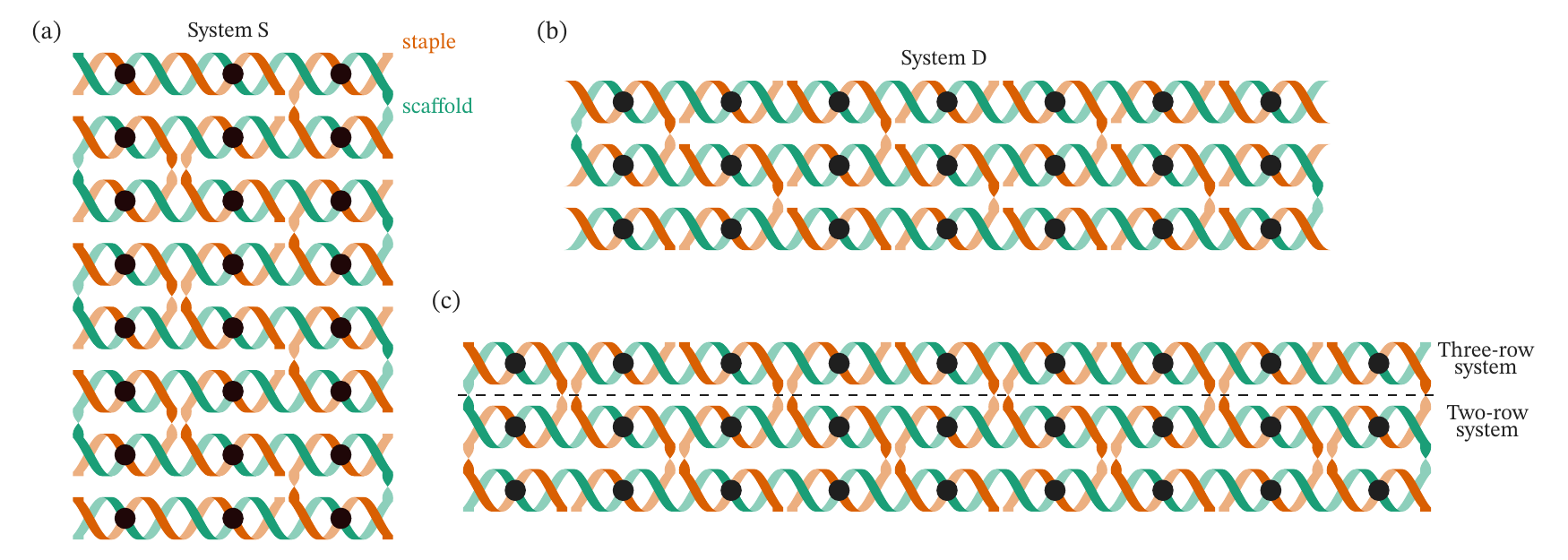}
    \caption{
        \label{fig:system-diagrams}
        Cartoon helix representations of the systems simulated in this study.
        Black circles identify the binding domains, which are both the fundamental design units of DNA origami and the level to which the model is coarse-grained.
        (a) System S, which has a 24-binding-domain scaffold and 12 two-binding-domain staple types.
        (b) System D, which has a 21-binding-domain scaffold and six two-binding-domain staple types, as well as eight single-binding-domain staple types.
        (c) Two- and three-row systems, with a dashed line showing the cut below which is the two-row system.
        The two-row system has an 18-binding-domain scaffold and nine two-binding domain staple types, while the three-row system has a 27-binding-domain scaffold and nine three-binding-domain staple types.
    }
\end{figure*}

However, we have previously developed a more coarse-grained model that represents DNA origami at the level of binding domains~\cite{cumberworth2018}.
A binding domain is the basic unit of origami design: in the final assembled state, each binding domain on the scaffold is bound to a complementary binding domain on a staple.
The model accounts for hybridization free energies, coaxial stacking of helices, and steric interactions.
To study nucleation behaviour more accurately, we have made some modifications to the model so that it better represents stacking and steric interactions and provides a more accurate representation of the chemical potential of the staples.
We provide details of the model and simulation methods in the \ac{SI}.

In this Letter, we use simulations with this coarse-grained model to show that nucleation can be a rate-limiting step in origami formation.
In order to be able to define free-energy barriers to nucleation, we must first define order parameters that can quantify the progress of self-assembly.
Here, we consider two order parameters: the numbers of (i) fully bound staples and (ii) bound-domain pairs.
The former effectively accounts for the size of the cluster and is analogous to the order parameter used in classical nucleation theory and DNA-brick self-assembly, while the latter provides us with a higher-resolution view of the mechanism by which staples bind.
By calculating the free energies associated with each possible value of the order parameter between assembled and unassembled states, we can determine whether barriers to assembly exist and, if so, estimate their magnitude.

To demonstrate the range of possible behaviours, we consider four systems: two that have been characterized in Reference~\citenum{cumberworth2018} (\cref{fig:system-diagrams}(a) and (b)) and two further systems (see below) with as many crossovers as possible (\cref{fig:system-diagrams}(c)).
The two previously studied designs are (a) system `S', a 24-binding-domain-scaffold system with 12 staple types, each with two binding domains (\cref{fig:system-diagrams}(a), which had been designed and simulated using the oxDNA model~\cite{snodin2016}; and (b) system `D', a 21-binding-domain-scaffold system with six two-binding-domain staple types and eight single-binding-domain staple types (\cref{fig:system-diagrams}(b)), which represents a subset of the system used by \citet{dannenberg2015} and \citet{dunn2015} 
In these two systems, each domain has a defined sequence.
We consider both sequence-specific and averaged interaction energies (see \ac{SI} for details).

\begin{figure*}[!ht]
    \centering
    \includegraphics{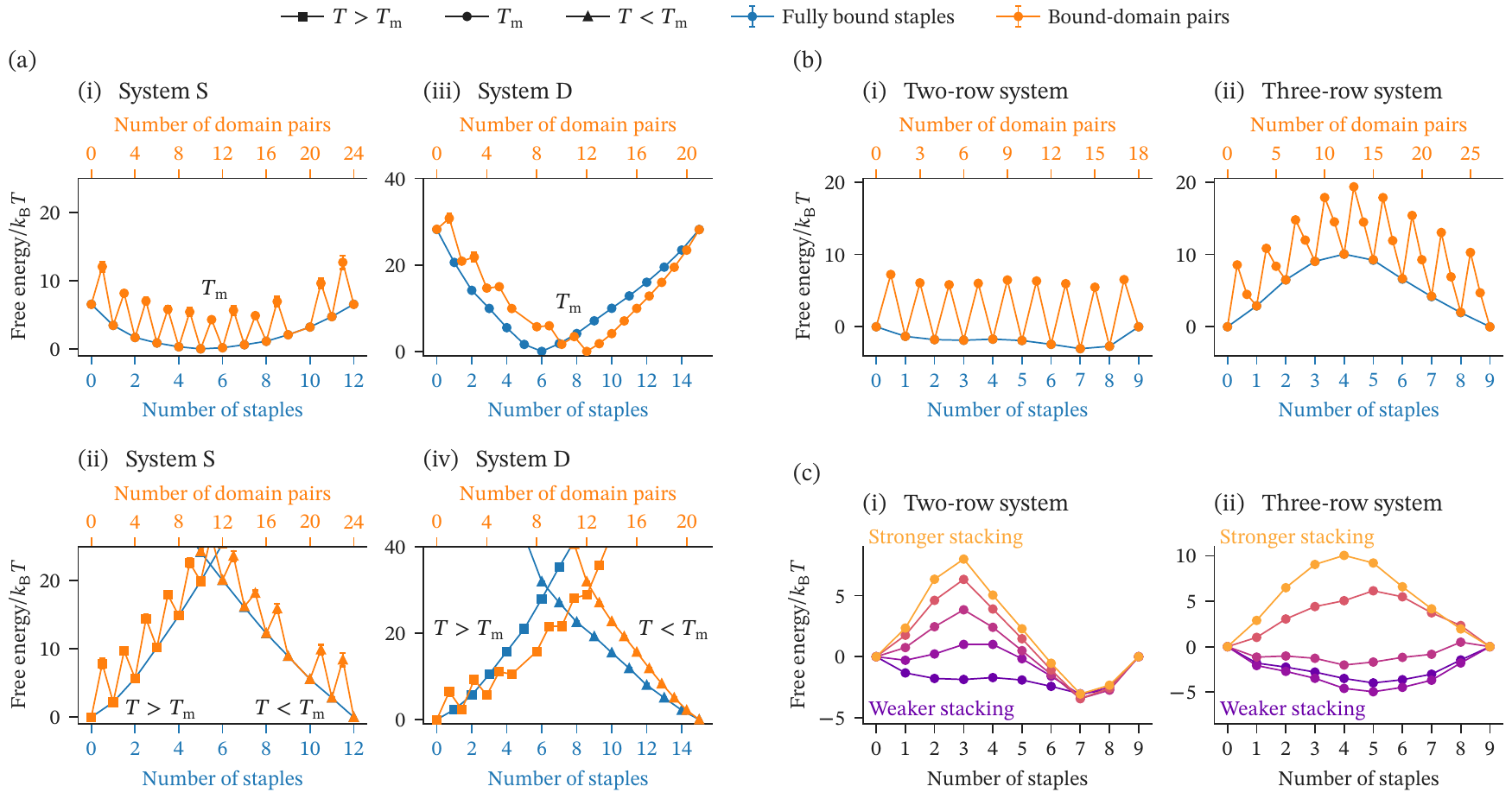}
    \caption{
        \label{fig:lfes}
        Free energies calculated for a range of values of selected order parameters.
        Here, both the number of fully bound staples and the number of bound-domain pairs are used as order parameters.
        (a) Free energies for system S, (i) and (ii), and system D, (iii) and (iv), at the melting temperature $T_\mathrm{m}$, (i) and (iii), and at both a temperature above and below $T_\mathrm{m}$, (ii) and (iv).
        The melting temperature is defined to be the temperature at which the free energy of the fully unassembled state is equal to the free energy of the fully assembled state.
        At the melting temperature for both system S (i) and system D (iii), the free energy is downhill to the favoured state along the number of fully bound staples, while along the number of bound domains, only small barriers related to fully binding each staple can be seen.
        Below and above the melting temperatures for system S (ii) and system D (iv), the free energies are again downhill to fully assembled and unassembled states, respectively.
        (b) Free energies for the two-row (i) and three-row (ii) systems at the melting temperature.
        In the two-row system, no nucleation barrier is observed, but the three-row system shows a clear nucleation barrier along both order parameters.
        Free energies at several temperatures for all systems are plotted in \cref{fig:lfes-temps}.
        (c) Free energies for the number of fully bound staples for the two-row (i) and three-row (ii) systems where the strength of the coaxial stacking parameter in the model has been varied.
        For the two-row system, a multiplier on the stacking parameter was increased from 1 to 2 in increments of 0.25.
        For the three-row-system, the multiplier on the stacking parameter was decreased from 1 to 0, also in increments of 0.25.
        Evidently, there is a strong dependence on the coaxial stacking of not only the magnitude, but even the presence of a nucleation barrier.
    }
\end{figure*}

\begin{figure}[!p]
    \centering
    \includegraphics{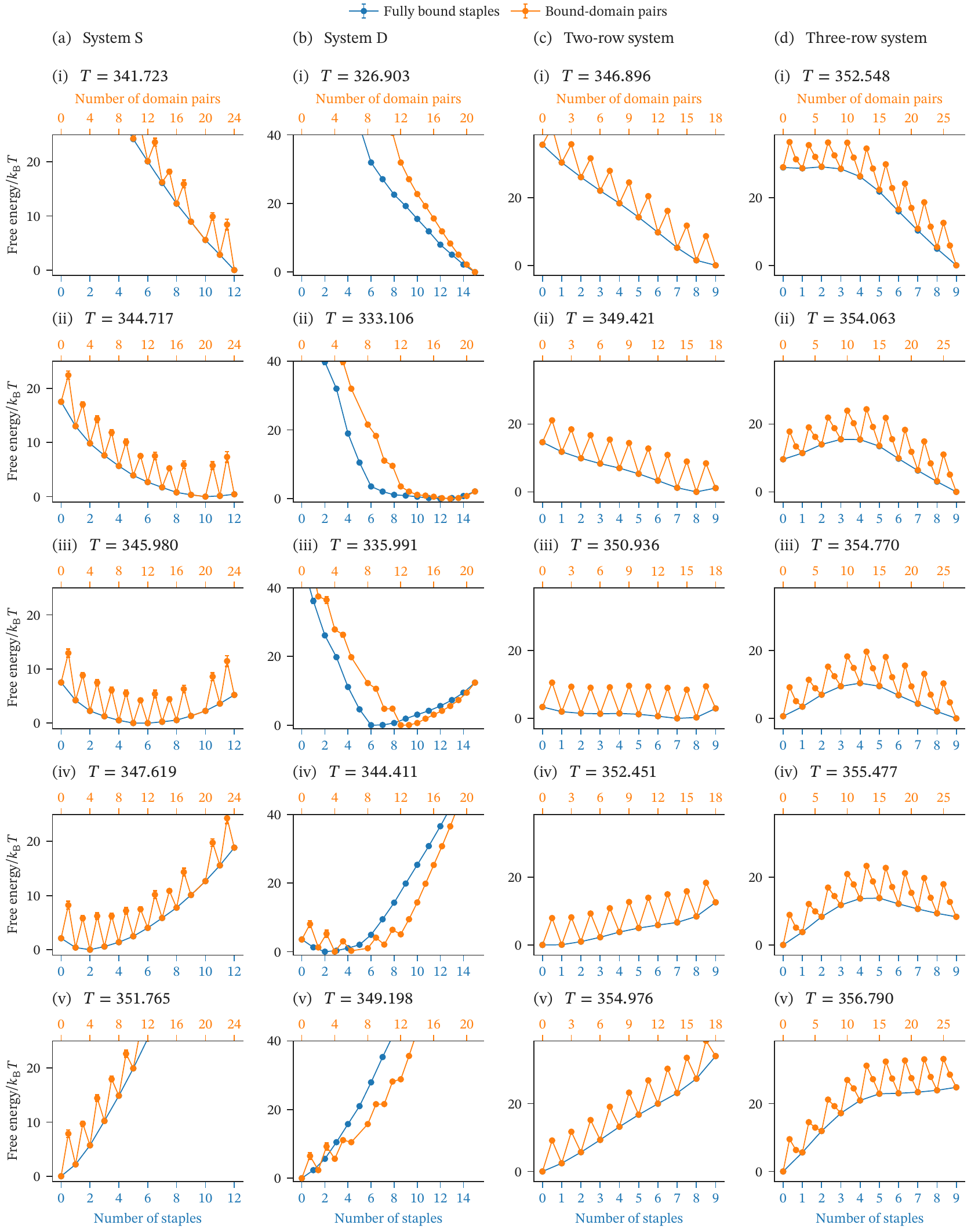}
    \caption{
        \label{fig:lfes-temps}
        Free energies for the number of fully bound staples and the number of bound-domain pairs at a range of temperatures below, near, and above the melting temperature.
        For systems S (a) and D (b), and the two-row system (c), the free energy along the number of fully bound staples is downhill to the favoured state, which shifts from assembled to unassembled as the temperature is lowered across several independent simulations.
        The three-row system (d) shows a barrier that appears near the melting temperature.
    }
\end{figure}

\begin{figure}[!p]
    \centering
    \includegraphics{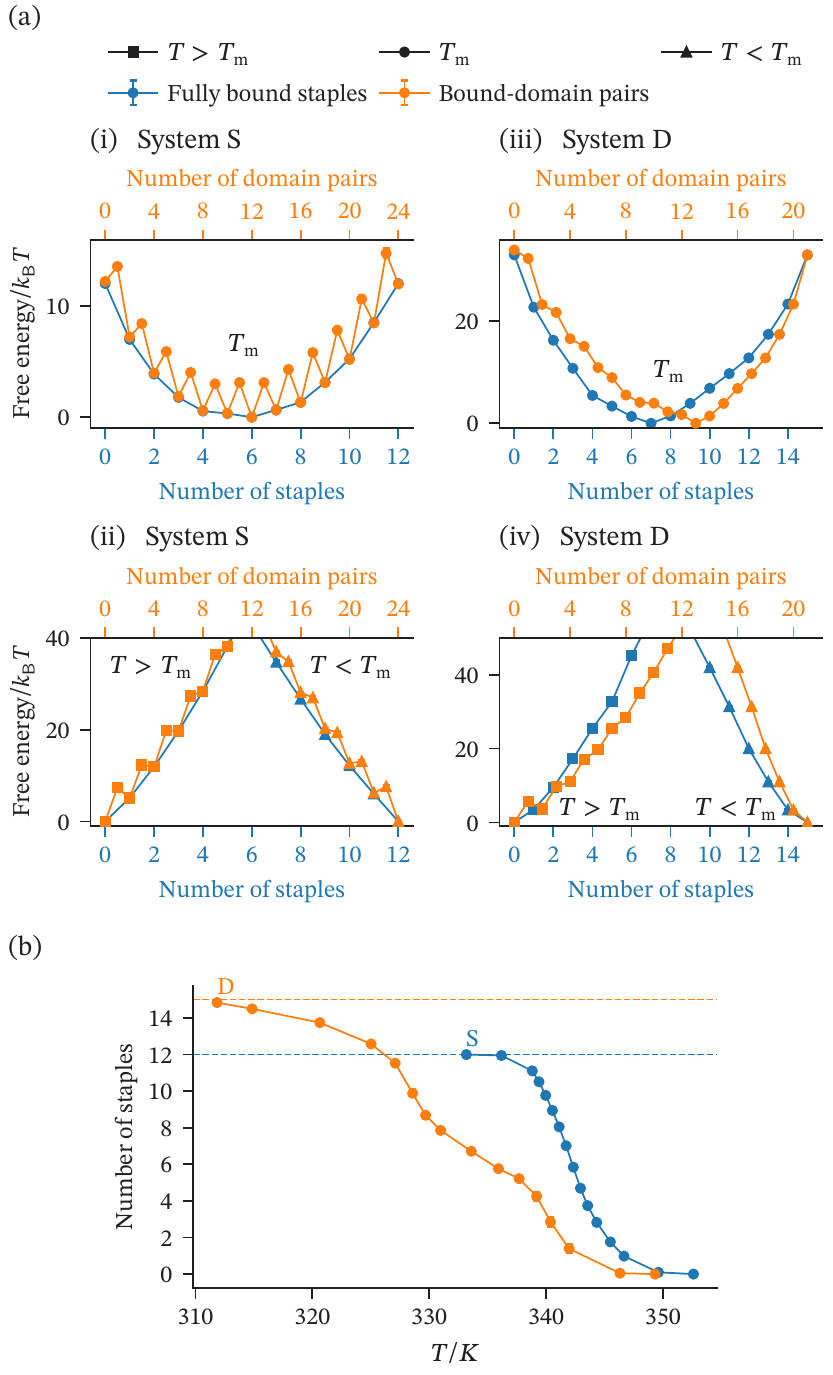}
    \caption{
        \label{fig:sequence-specific}
        Free energies and expectation values using sequence-specific hybridization free energies.
        (a) Free energies for the number of fully bound staples and the number of bound-domain pairs for system S, (i) and (ii), and system D, (iii) and (iv), at the melting temperature $T_\mathrm{m}$, (i) and (iii), and at both a temperature above and below $T_\mathrm{m}$, (ii) and (iv).
        The melting temperature is defined to be the temperature at which the free energy of the fully unassembled state is equal to the free energy of the fully assembled state.
        (b) The expectation values of the number of fully bound staples as a function of the temperature.
        The overall qualitative behaviour is the same as when averaged hybridization free energies are used.
    }
\end{figure}

The free energies for systems S and D show no nucleation barrier along either order parameter considered with both averaged hybridization free energies (\namecrefs{fig:lfes} \labelcref{fig:lfes}(a) and \labelcref{fig:lfes-temps}) and sequence-specific hybridization free energies (\cref{fig:sequence-specific}).
For computational simplicity, we define the melting temperature as the temperature at which the free energies of the fully assembled and fully unassembled states are equal.
For both systems, at high (low) temperatures, the unassembled (assembled) state is favoured, but at the melting temperature, the free energy as a function of the number of fully bound staples is lowest for the partially assembled state.
In \cref{fig:lfes}(a)(i) and (ii), the free energies along the number of bound-domain pairs alternate between higher and lower values; this is consistent with the second binding domain of a staple having a lower entropic cost of binding than the first binding domain of a staple, and with a small easily surmountable barrier for staples that are near their individual melting points.

Although there is no nucleation barrier in these specific systems, it is known from experiments that hysteresis sometimes arises in DNA-origami systems.
To determine conditions under which a nucleation barrier can arise, we first note that, in the context of DNA-brick self-assembly, it was shown that increasing the coordination number of the assembly units increases the barrier height~\cite{reinhardt2016b}.
Typical DNA bricks have a coordination number of four, while for the DNA origami designs of systems S and D, it is two at most.
To test whether the same principle might apply in the context of DNA origamis, we increase the number of binding domains per staple as a way of increasing the coordination number.
To this end, we design a set of systems that have the maximum number of crossovers possible for a system with a given number of staple types and helices in the assembled structure (\cref{fig:system-diagrams}(c)).
In the assembled state of these designs, the scaffold forms a series of rows in a single plane, each of which comprises a single helix.
In each column, a single staple crosses over all helices formed by the scaffold, and thus, the number of binding domains per staple corresponds to the number of rows in the design.
Because we are more interested in trends for these systems, and because there was no qualitative difference in the results between the sequence-specific and averaged hybridization free energies for systems S and D, we consider only the averaged hybridization free energies for our designed systems.
In this study, we restrict ourselves to systems that had nine binding domains per row and consider two- and three-row variants.

\begin{figure*}[!ht]
    \centering
    \includegraphics{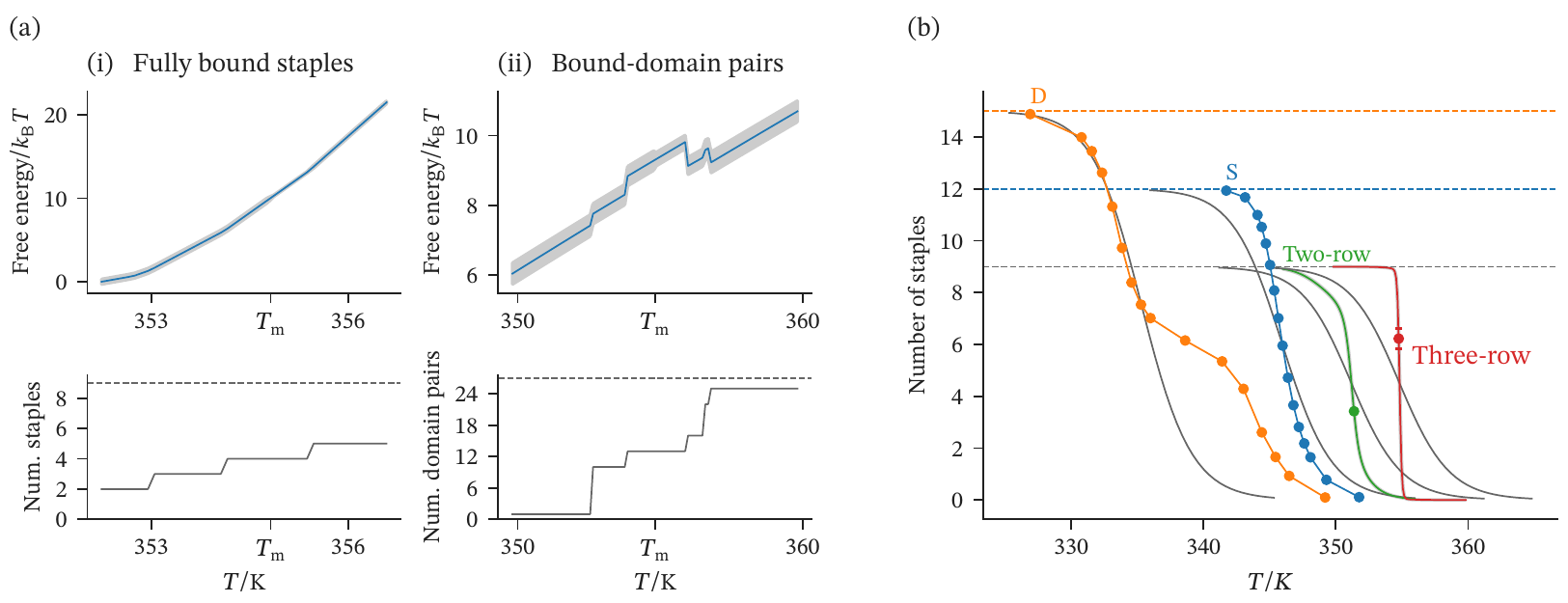}
    \caption{
        \label{fig:means}
        (a) Barrier height as a function of temperature for the three-row system, with the position of the peak plotted below.
        The plots in (i) include the entire domain over which the barrier along the number of fully bound staples is defined; outside of these temperatures, the free energies are either monotonically increasing or decreasing (\cref{fig:lfes-temps}(d)).
        The barrier height for the number of bound-domain pairs is calculated by taking the difference between the value at the peak and the value at the local minimum; in all cases, the local minimum is located at $N_\mathrm{BD}^* - 1$, where $N_\mathrm{BD}^*$ is the number of bound-domain pairs at the peak (\namecrefs{fig:lfes} \labelcref{fig:lfes}(a)(iv) and \labelcref{fig:lfes-temps}(d)).
        The nucleation barrier can be seen to disappear a few degrees below the melting temperature.
        (b) Expectation values of the number of fully bound staples as a function of the temperature.
        The grey lines centred on the two- and three-row system are the curves that result from assuming the binding domains act independently (see \ac{SI} for calculation details).
        The points show temperatures used in the simulations; for the two- and three-row systems, there is only one point as they were simulated with \ac{US}.
        The lines between the points for system S and system D are drawn only to guide the eye, while the lines for the two- and three-row system are calculated via extrapolation (see \ac{SI} for details).
        The light grey around the extrapolated lines represents the uncertainty in the extrapolated values.
        In both (a) and (b), the dashed lines indicate the value of the order parameter at the fully assembled state for the system (with the corresponding colour in (b)).
        The three-row system shows an unusually sharp transition between unassembled and assembled states.
    }
\end{figure*}

\begin{figure}[!p]
    \centering
    \includegraphics{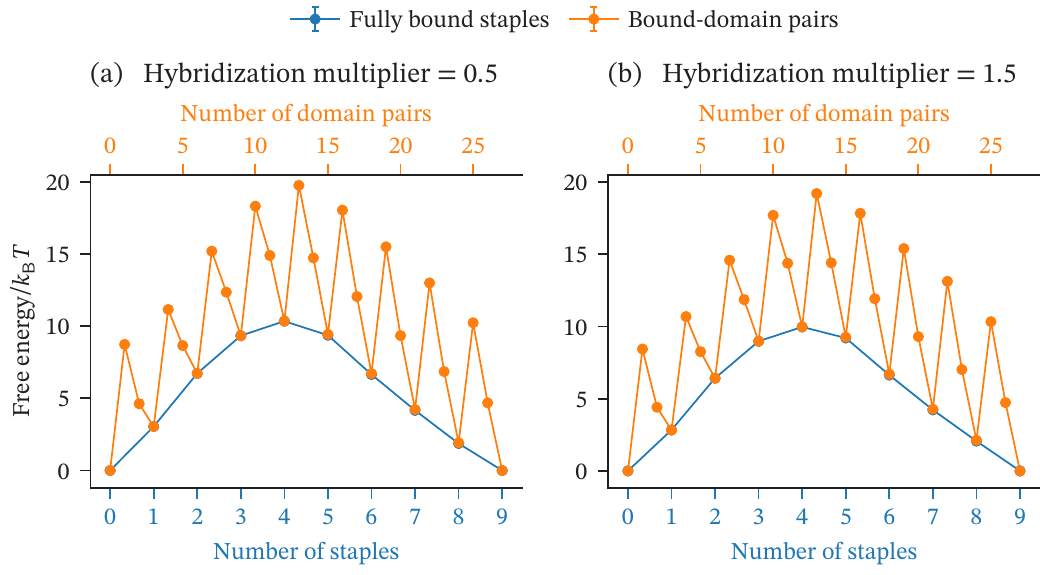}
    \caption{
        \label{fig:hyb-mult}
        Free energies for the number of fully bound staples and the number of bound-domain pairs with two different multipliers applied to the hybridization free energies for the three-row system.
        The free energies were calculated by extrapolation with the \ac{MBAR} method from the simulations run with no multiplier.
        The multiplier is applied to the average enthalpies and entropies of binding and misbinding for each binding domain.
        The barrier is nearly independent of the strength of the hybridization free energy when all binding domains are shifted by the same amount.
    }
\end{figure}

The free energy for the two-row system along the number of fully bound staples (\cref{fig:lfes}(b)(i)) at the melting temperature has no nucleation barrier.
By contrast, the three-row system has a clear barrier to assembly, the maximum of which occurs nearly halfway along to the fully assembled state, at four fully bound staples, with a magnitude of $\sim$\SI{10}{\kBT} (\cref{fig:lfes}(b)(ii)).
In both systems along the free energies of the number of bound-domain pairs, similar to the results seen for systems S and D, there are peaks at regular intervals, occurring with a frequency equal to the number of binding domains in the staple.
In the three-row system, these peaks effectively add on to the barrier seen in the number of fully bound staples, giving a total barrier of around \SI{20}{\kBT}.
The true free-energy barriers to self-assembly are likely somewhat higher than this due to the initial binding of the first nucleotide of a domain; however, a higher-resolution model would be needed to determine their magnitudes.
As the temperature is lowered to below the melting point, the barrier along the number of fully bound staples disappears after a few degrees, and the barrier along the number of bound-domain pairs also decreases substantially (\cref{fig:means}(a)).
On the other hand, using average hybridization free energies that are 50\% smaller or larger, while substantially shifting the melting temperature, had almost no effect on the barrier height (\cref{fig:hyb-mult}): although they bind less (more) strongly, the entropic cost of binding is lower (higher) because of the shift in melting temperature, and the two effects appear to cancel each other.

Since all binding domains by construction have the same hybridization free energy, we might expect the systems to assemble over a relatively narrow temperature range.
However, the ranges within which the S, two-row, and three-row systems transition are narrower than they would be for the same number of independent binding domains (\cref{fig:means}(b)).
The three-row system displays an especially sharp transition, from entirely unbound to entirely bound in less than $\sim$\SI{1}{\kelvin}.
The observed narrowing of the assembly as a function of the temperature in all studied systems implies that cooperativity is involved in the assembly process, but the nucleation barrier observed in the three-row system implies not only stronger cooperativity, but also the presence of a particular type of cooperativity.
By investigating the origins of this cooperativity further, we may therefore be able to determine under what conditions nucleation barriers are likely to occur in DNA-origami self-assembly.

Cooperative behaviour of staples and binding domains can occur via three routes: closing of scaffold loops, initial binding of the domain of a staple to the scaffold, and coaxial stacking of binding domains adjacent in the same helix~\cite{majikes2021}.
The first route, the closing of loops, could plausibly lead to a nucleation barrier, but to be a viable pathway, it would generally require initial staples to bind more strongly than those that bind once loop closure becomes thermodynamically favourable.
Since we use averaged hybridization free energies, this mechanism cannot dominate in this case.
The second route leads to the jaggedness of the free energies along the number of bound-domain pairs in \cref{fig:lfes}(a), but it cannot explain the barrier we observe along the number of fully bound staples.
We therefore focus our investigation on the stacking of adjacent binding domains along the same helix.

When a fluctuation occurs in a system so that several staples bind concurrently in such a way that they can stack with each other, the energetic gain can be sufficient to overcome the entropic cost of binding at a temperature that is higher than it would be for a given staple in isolation.
The stronger the stacking per staple, whether by a more favourable stacking energy at each domain or by having more domains to stack per staple, the higher the temperature at which a cluster of staples is able to bind relative to the staples in isolation.
This increased temperature difference also leads to a higher barrier, as the fluctuation needed for a given staple to bind has a higher entropic cost.
We therefore anticipate that the more favourable the stacking energy, the greater the cooperativity and the larger the nucleation barrier will be.

\begin{figure}[!p]
    \centering
    \includegraphics{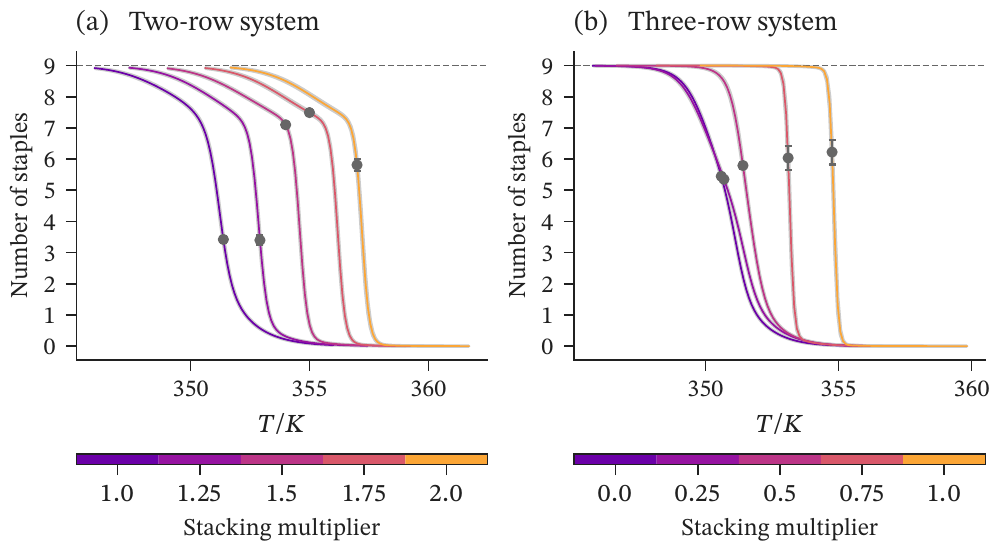}
    \caption{
        \label{fig:means-stacking}
        The expectation values of the number of fully bound staples as a function of the temperature for a range of stacking energies.
        The light grey around the extrapolated lines represent the uncertainty in the extrapolated values.
        In both (a) and (b), the dashed lines indicate the value of the order parameter at the fully assembled state for the system.
        The three-row system shows an unusually sharp transition between the assembled and unassembled states, but this is reduced substantially with lower stacking energies.
    }
\end{figure}

To test this hypothesis, we run simulations where we vary the stacking energy parameter.
The free energies in \cref{fig:lfes}(c) reveal that halving the stacking energy in the three-row system leads to the complete disappearance of the barrier.
Moreover, the temperature range of the transition broadens as the stacking energy is reduced (\cref{fig:means-stacking}).
On the other hand, in the two-row system, a clear barrier is seen as the stacking energy is scaled by 1.5 or more (\cref{fig:lfes}(c)).

\begin{figure}[!t]
    \centering
    \includegraphics{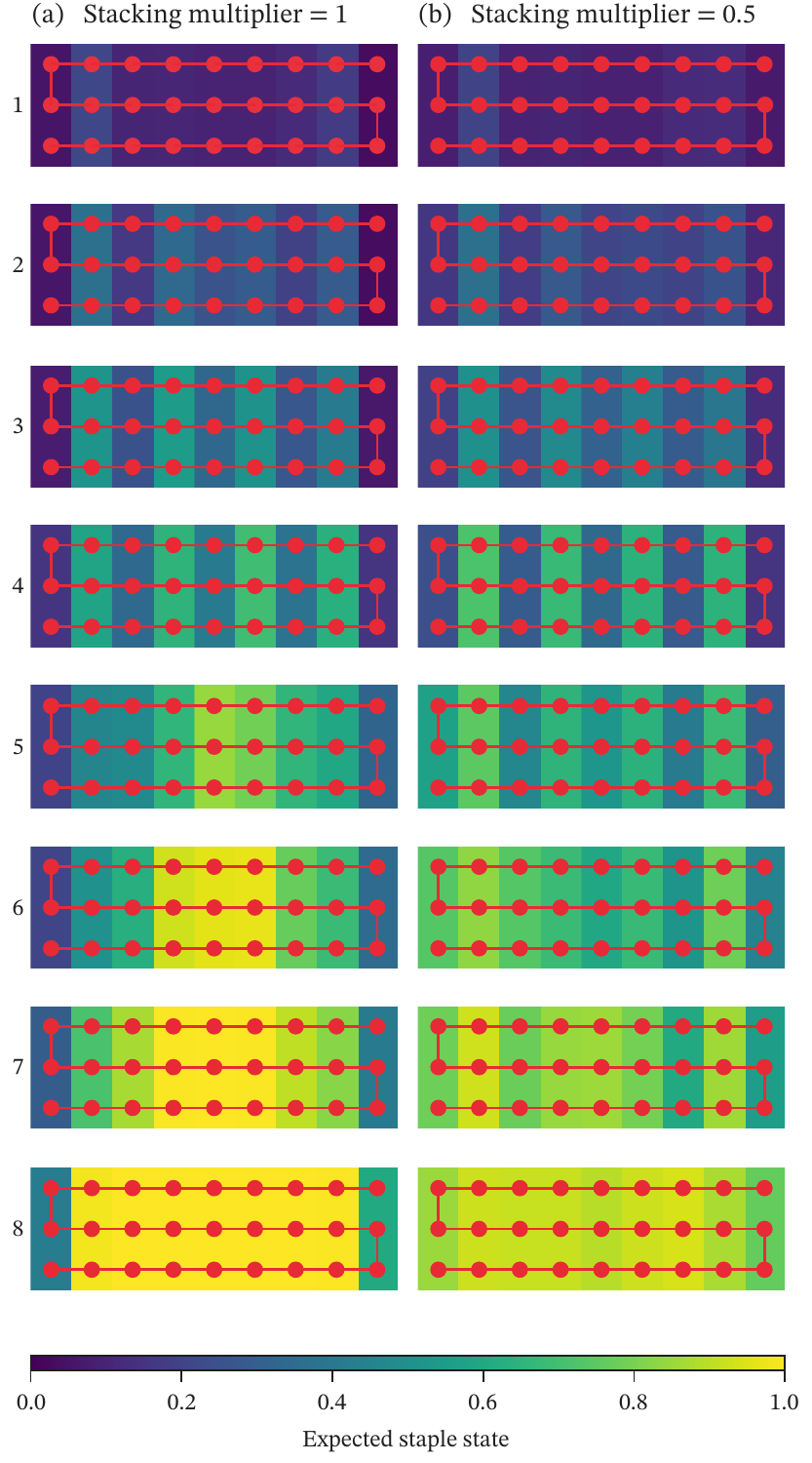}
    \caption{
        \label{fig:freqs}
        Expectation values of the staple state for each staple type at the melting temperature in the three-row system plotted as heat maps.
        For a given total number of fully bound staples, the heat maps show the fraction of configurations that have a staple type fully bound.
        The number of fully bound staples used for each set of expectation values is given to the left of the heat maps in each row.
        A diagram of the scaffold of the design is superimposed on each heat map.
        In (a), the stacking energy is set to the model's standard value~\cite{cumberworth2018}, while in (b) it is set to half that value.
        With full stacking, the assembly pathway indicates that nucleation tends to begin in the middle of what will become the assembled state and then grows outward; with half stacking, staples bind uniformly to the scaffold during assembly.
    }
\end{figure}

\begin{figure}[!p]
    \centering
    \includegraphics{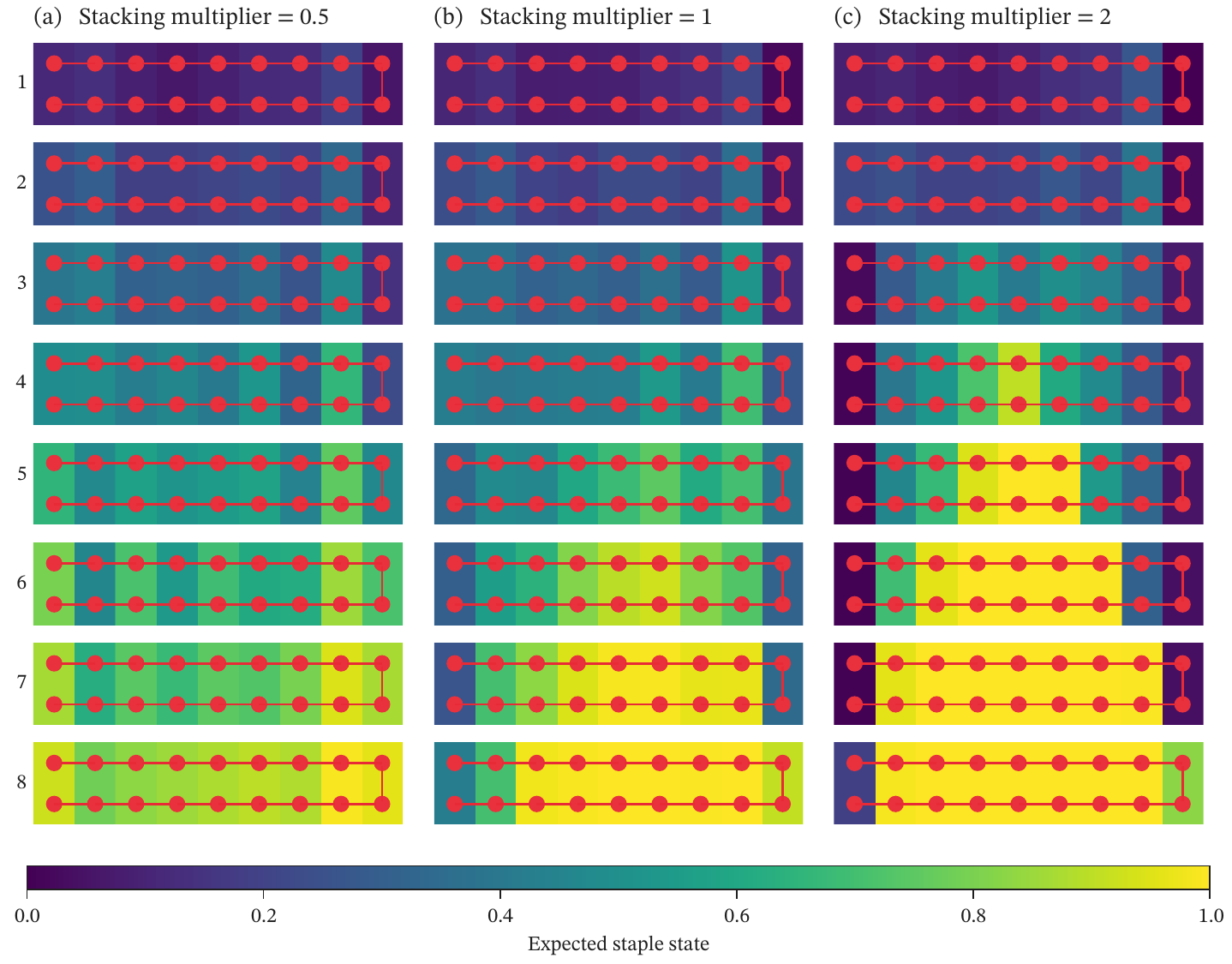}
    \caption{
        \label{fig:freqs-two}
        Expectation values of the staple state for each staple type at the melting temperature in the two-row system plotted as heat maps.
        For a given total number of fully bound staples, the heat maps show the fraction of configurations that have a staple type fully bound.
        The number of fully bound staples used for each set of expectation values is given to the left of the heat maps in each row.
        A diagram of the scaffold of the design is superimposed on each heat map.
        In (a), (b), and (c), the stacking energy is set to half, equal to, and double the model's standard value~\cite{cumberworth2018_sup}, respectively.
        With double the stacking energy, the system shows a clear pattern of nucleation, where the initial staple tends to bind in the middle of what will become the assembled structure, with subsequent growth outwards.
        While \cref{fig:lfes}(b)(i) shows that the two-row system has no nucleation barrier at the standard stacking energy, it still shows a tendency to bind in the middle and grow outwards, which indicates that it is close to having a nucleation barrier.
        This is demonstrated by a small barrier appearing in \cref{fig:lfes}(c)(i) even with a stacking energy multiplier of 1.25.
        At half the stacking energy, the staples bind relatively uniformly to the scaffold.
    }
\end{figure}

We investigate the associated change in the assembly pathway by calculating expectations of individual staple states for a given number of fully bound staples.
In \cref{fig:freqs}(a), we show that with the full stacking energy in the three-row system, after the barrier peak, there is a higher density of bound staples at the centre, which becomes more intense and spreads outward as the number of fully bound staples increases.
With half the original stacking energy, no such cluster appears (\cref{fig:freqs}(b)).
A similar pattern is seen with the two-row system when comparing simulations with multipliers on the stacking energy of 0.5, 1, and 1.5  (\cref{fig:freqs-two}).
These results indicate that a nucleation barrier and assembly pathway can be designed either by making the stacking energy more favourable (for example, by changing the salt concentration, by modifying the sequence pairings that occur at breakpoints, or even by using modified nucleobases which have different stacking interactions) or by increasing the number of stacking interactions in the origami design.

In summary, we have demonstrated that nucleation barriers in DNA origami depend on the coaxial stacking between helices and that some designs have no barrier at all.
Small or non-existent barriers and the consequent reversibility in the transition may be useful in a number of applications, since origamis may be switched between assembled and unassembled states by changing solution conditions for functional purposes.
We have also shown that origamis that do exhibit nucleation barriers can be designed by maximizing the number of crossovers in a system, thus increasing the effective coordination number, which results in a high degree of cooperativity and which in turn can be tuned by modifying the number of binding domains per staple.
Since the resulting nucleation barriers are still surmountable, but the temperature range over which a transition occurs is very narrow, one can envisage applications such as molecular-scale thermometers or, by suitable functionalization, other molecular sensors.

Our results provide a rationalization for both the success of isothermal assembly and the hysteresis sometimes observed in temperature-ramp protocols: origami designs either have no barrier, or one that only exists around the melting temperature.
Moreover, our results suggest that systems can be designed with barriers optimized for both assembly time and yield in an isothermal assembly protocol.
If staples bind to multiple places on the scaffold concurrently, the rearrangement times of the helices in different partially assembled chunks could be very slow, potentially leading to jammed states.
A barrier allows for assembly pathways that begin locally and then grow out from that point.
While the barrier observed here disappears a few degrees below the melting temperature, the assembly temperature could be tuned to be just below the melting temperature to retain the barrier and still have a good yield due to the sharp transition.

One possible difference between the self-assembly behaviour of DNA origami and DNA bricks is the latter's propensity for aggregating in such a way as to prevent full assembly.
In studies of DNA bricks, it was found that at lower temperatures, incidental interactions led to the aggregation of partially assembled structures, creating a rugged free-energy landscape that inhibits the assembly process~\cite{reinhardt2014,jacobs2015,sajfutdinow2018,zhang2020}.
Our approach cannot be directly used to simulate such aggregation in DNA-origami systems because it does not include free staples or other scaffolds; however, since the free energies along the number of bound-domain pairs are always downhill after the binding of the first domain of a staple, this would seem to imply that the staples tend to bind fully and have fewer unhybridized segments available.
This makes DNA origami less prone to aggregation, as the partially assembled structures have fewer possibilities for incidental interactions with each other.
This observation explains why isothermal assembly below the melting temperature can so often be successful in DNA-origami self-assembly.

Here we have focused on averaged hybridization free energies, but increased heterogeneity in the individual staple hybridization free energies could lead to lower barriers.
With sufficiently disparate staple melting temperatures, the stacking energy would be insufficient to allow multiple staples to bind in such a way that they stack with each other to overcome the entropic cost of binding.
If a nucleation barrier is desired, then it may therefore prove helpful to design staple sequences that have interaction energies that are as monodisperse as possible.
Similar considerations have been shown to hold for DNA bricks~\cite{jacobs2015}, although the aim in that case is usually to reduce the nucleation barrier.

In order to be able to probe the thermodynamics of DNA-origami self-assembly, we used a coarse-grained model and relatively small system sizes to ensure sufficiently rapid convergence.
Although larger DNA-origami structures with complex scaffold routing might be subject to other kinds of free-energy barriers to self-assembly, many commonly used origami designs are scaled-up versions of the systems we have considered, and given that the barrier height scales with the per-staple stacking strength, rather than a global measure of the origami size, we expect our key findings to apply to such systems.
Moreover, with the recent development~\cite{pound2009,said2013,erkelenz2014,brown2015,nafisi2018,engelhardt2019,bush2020} of scaffolds shorter than the M13mp18 phagemid often used in origami designs, we speculate that the use of smaller scaffolds may become more popular, including scaffolds that enable highly cooperative maximum-crossover designs with monodisperse hybridization free energies.

One key message is that our results reveal that nucleation in DNA-origami self-assembly is fundamentally different from the nucleation behaviour of DNA bricks, and that it is possible to control, and even eliminate, the size of the barrier by judicious staple design.
Such design would provide a tool for optimizing assembly times and yields and for tailoring origamis to specific functional applications.

    \section*{Data Availability Statement}
    The data underlying this study are openly available in Zenodo at \href{https://doi.org/10.5281/zenodo.6414264}{https://doi.org/10.5281/zenodo.6414264}~\cite{replication-package}.

    \begin{acknowledgments}
        We would like to thank Rosana Collepardo-Guevara and Thomas Ouldridge for their insightful comments on an early version of this manuscript.

    \end{acknowledgments}
    \bibliography{tex/main_arxiv}
    \onecolumngrid
    \clearpage
    \pagebreak

    \twocolumngrid
    \title{\textcolor{textblack}{Supporting information\texorpdfstring{:\\}{:} Simulations of DNA-origami self-assembly reveal design-dependent nucleation barriers}}
    \maketitle
    \supplemental
    \begin{bibunit}[apsrev4-2]
    \makeatletter\write\@bibunitaux{\string\citation{REVTEX42Control}}\makeatother
    \makeatletter\write\@bibunitaux{\string\citation{apsrev42Control}}\makeatother
    \acresetall

\begin{acronym}
    \acro{bp}{base pair}
    \acro{CB}{configurational bias}
    \acro{CT}{conserved topology}
    \acro{CTRG}{conserved-topology recoil-growth}
    \acro{LFE}{Landau free energy}
    \acroplural{LFE}[LFEs]{Landau free energies}
    \acro{MBAR}{multistate Bennett acceptance ratio}
    \acro{MC}{Monte Carlo}
    \acro{NN}{nearest-neighbour}
    \acro{nt}{nucleotide}
    \acro{REMC}{replica-exchange Monte Carlo}
    \acro{RG}{recoil-growth}
    \acro{SI}{Supporting Information}
    \acro{US}{umbrella sampling}
    \acro{MWUS}{multi-window umbrella sampling}
    \acro{REMWUS}{replica-exchange multi-window umbrella sampling}
    \acro{WHAM}{weighted-histogram analysis method}
\end{acronym}

    \noindent In this Supporting Information, we give a description of the DNA-origami lattice model, which includes improvements over and clarifications to the version published previously~\cite{cumberworth2018_sup}.
We also describe the simulation and analysis methods employed by this study.
Finally, additional supporting figures are included at the end.

The simulation, analysis and plotting code, as well as all input scripts and intermediate calculation results, are available as a replication package~\cite{replication-package_sup}.

\section{Model description}

\subsection{State space}
\label{sec:state-space}
\begin{figure}
    \centering
    \includegraphics{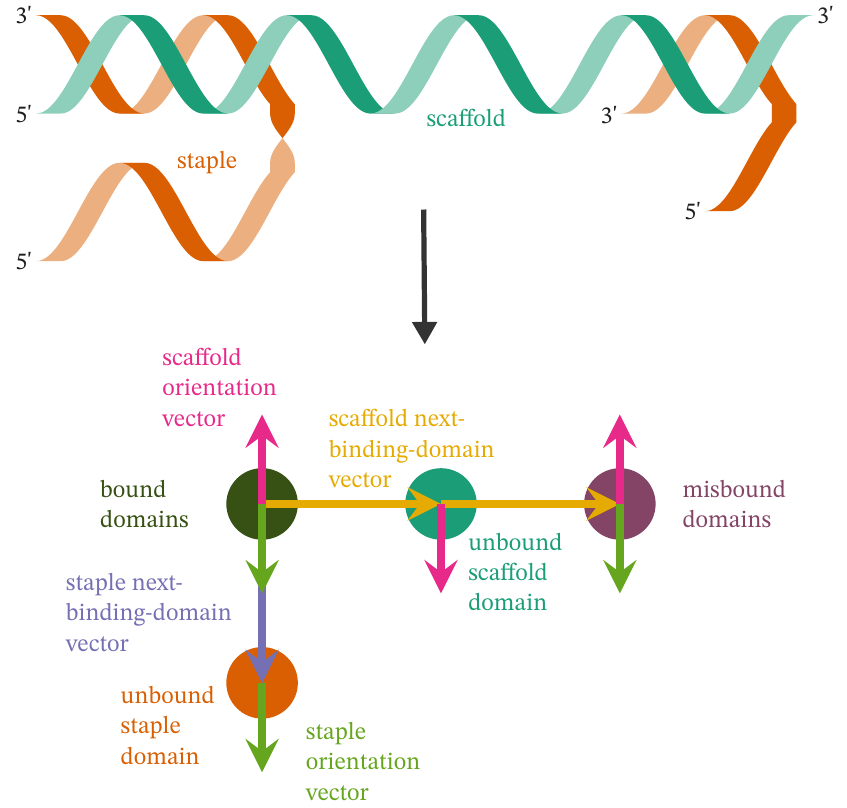}
    \caption{
        \label{fig:model-overview}
        Schematic illustrations of the basic elements of the model.
        A cartoon helix representation is given on the top, and on the bottom is a representation of the same configuration in our model.
        There is one scaffold with three binding domains, one staple with two binding domains, and one staple with a single binding domain.
    }
\end{figure}

\begin{figure*}
    \centering
    \includegraphics{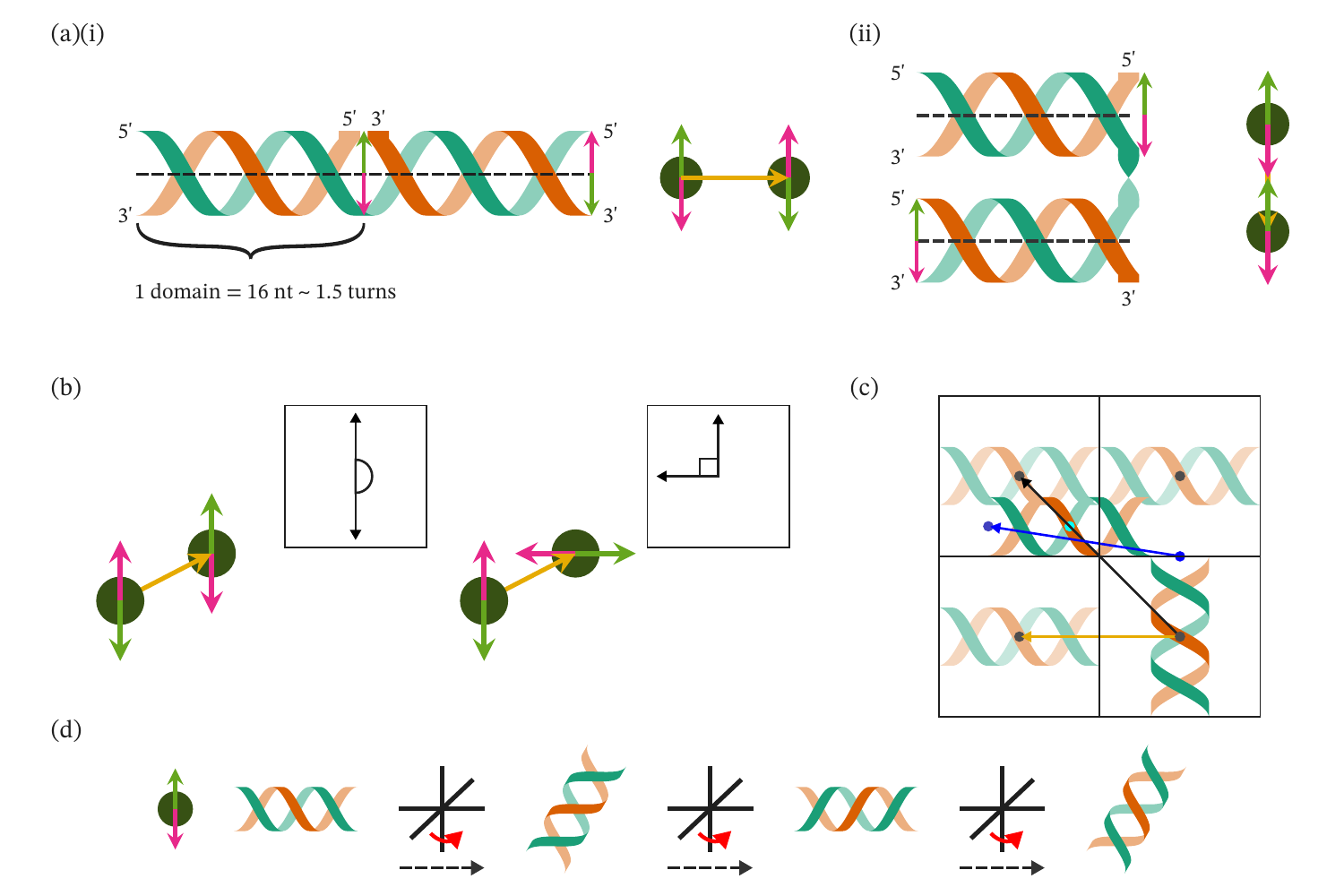}
    \caption{
        \label{fig:helical-twist}
        Representation of helices in the model.
        (a) Two 16-\ac{nt} binding domains, which in B-form DNA corresponds to about 1.5 turns of the helix.
        In (i), both binding domains are part of the same helix, while in (ii), they are part of separate helices with a kink between them.
        (b) Orientation vectors and helical phase.
        The boxes are projections of the scaffold orientation vectors of the two binding domains onto a plane normal to the next-binding-domain vector, with the dihedral angle indicated.
        Left: the orientation vectors are consistent with two stacked 16-\ac{nt} (1.5 turn) binding domains.
        Right: the orientation vectors are consistent with two stacked 8-\ac{nt} (0.75 turn) binding domains.
        (c) Physical interpretation of the next-binding-domain vector when considering two orthogonal helices.
        The helix of the first binding domain is drawn on the bottom right lattice site, while the helix of the second binding domain is drawn after rotation without constraining it to be on an adjacent lattice site, as well as drawing it centred on the remaining three lattice sites (with lighter colours).
        Black points are placed at the centre of each lattice site.
        The black vector points from the centre of the first binding domain to the centre of the top left lattice site, which happens to pass through the centre of the second binding domain (marked by a teal point).
        The blue vector points from the end of the first binding domain to the end of the second binding domain.
        The yellow vector is the result of coarse-graining the blue vector to the next-binding-domain vector of our model.
        (d) A single pair of bound domains only constrains the helical axis to a plane.
    }
\end{figure*}

The basic elements of state space are laid out in the main text and in the original publication of the model~\cite{cumberworth2018_sup}, the latter of which also provides the motivation for these choices.
Here, we summarize these elements and provide a number of clarifications.
The basic units of DNA origami design, binding domains, are represented as particles on a simple cubic lattice.
We will occasionally refer to binding domains as simply `domains'.
Contiguous binding domains on a given chain are constrained to occupy adjacent lattice sites.
For each binding domain on the scaffold, there is a complementary binding domain on a staple that it is intended to bind to.
Lattice sites can have an occupancy of zero (unoccupied), one (unbound) or two (bound or misbound), where the number indicates how many binding domains are present at that site (\cref{fig:model-overview}).
Every binding domain also has an orientation vector $\uvec{o}_i$.
The vector that points from one binding domain to the next on the same chain is referred to as the `next-binding-domain vector' $\uvec{n}_i$, which does not necessarily coincide with the helical axis in a bound state.
When two domains are bound or misbound, the orientation vectors must sum to zero.

The definition of a binding domain in the model may differ from the definition of a binding domain for a given design, as here the binding domains must be all approximately the same length, or number of residues.
If a design has binding domains whose lengths are integer multiples of each other, the binding domain as defined in our model will be the smallest binding domain in the design, with larger binding domains in the design being represented by multiple binding domains in the model.
Designs commonly vary the length of the binding domains by one or two residues to reduce internal stresses in the final structure: these may be represented in our model as a single binding domain since the total change in length is relatively small.
However, if the binding domains differ substantially in length and are not integer multiples of each other, it may not be feasible to represent the design with our model.

In states involving two domains bound to each other (a bound-domain pair), the orientation vector of the model maps to a vector that points out orthogonally from the helical axis to the position of the strand at the end of the helix in the current binding domain (\cref{fig:helical-twist}(a)).
In the case of a scaffold chain, the positive direction is defined as $5'$ to $3'$, while in the case of a staple chain, it is defined as $3'$ to $5'$.
From this mapping, we can identify configurations of two bound-domain pairs, where at least two of the binding domains are contiguous on a chain, as forming a single helix if the dihedral angle between the planes defined by the orientation vectors and next-binding-domain vector is consistent with the number of turns of the helix per bound-domain pair (\cref{fig:helical-twist}(b)).
This identity is used in the potential energy function to determine steric penalties and stacking interactions.

With no explicit helical-axis vector in the model, a single bound-domain pair will only implicitly define the helical axis to lie within a plane (\cref{fig:helical-twist}(d)).
The helical axis is not resolved until an adjacent domain enters a bound state in the same helix.
If a binding domain contiguous to one of the two bound domains enters a bound state that is not in the same helix, then the helical axes of the first bound pair and the new bound pair will become more restricted in a configuration dependent way, but will not be fully resolved.
This is accounted for with the potential energy function, which will be discussed in the following subsection.

Because we are restricted to a simple cubic lattice, all binding domains that may be modelled fall into four classes, based on the dihedral angle prescribed by the number of turns.
These classes differ by the fraction of a turn that remains after the $n$ whole turns that make up the double helix for a binding domain in a hybridized state.
We refer to these as whole-turn binding domains (remainder of zero), quarter-turn binding domains, half-turn binding domains, and three-quarter-turn binding domains.
However, whole-turn binding domains are not useful in origami design, at least using our definition of binding domains, as they do not allow for crossovers between parallel helices, so we do not consider them further.

The mapping between the model geometry and more detailed representations is not intended to be exact.
There are two related issues that should be mentioned here.
First, parallel helices with crossovers between them are represented as being on adjacent lattice sites, yet the distances between the centres of helices along the same helix and across parallel helices will in general not be the same.
For small binding domains, these distances are approximately the same, but for longer binding domains, the approximation will become progressively worse.
We mostly consider 8-\ac{nt} and 16-\ac{nt} binding domains with our model; an 8-\ac{bp} bound-domain pair is about \SI{2.7}{\nano\meter} long, assuming B-form DNA geometry, while a 16-\ac{bp} is about \SI{5.5}{\nano\meter} long.
Assuming approximately \SI{1}{\nano\meter} spacing between parallel helices in an origami structure~\cite{rothemund2006_sup}, we get a distance between the centres of parallel helices that is between these two values, \SI{3}{\nano\meter}.

The second mapping issue involves the physical meaning of the next-binding-domain vector.
A simple way to visualize this vector is to have it point from the centre of the first binding domain to the centre of the second binding domain.
While this is sufficient for unbound domains and binding domains that form a single helix, the mapping does not always work for bound domains that are not in the same helix.
For two orthogonal helices, the vector would point approximately towards a lattice site diagonal to the current site, yet we only allow contiguous binding domains to be on adjacent lattice sites.
To resolve this, consider that the aim is to bin configurations represented at a higher level of detail to our lattice model in a consistent way, and to broadly capture the geometry of assembled and partially assembled states.
The assembled structures able to be considered with this model do not involve these configurations, and the exact geometry of the partially assembled states is unlikely to be critical for our purposes.
Thus, when binning configurations with orthogonal helices, the next-binding-domain vector is defined to point from the end of the first helix to the end of the second helix (\cref{fig:helical-twist}(c)).
Finally, for two contiguous bound domains that are in separate parallel helices (i.e.~they have a crossover between them), it is sufficient to consider the next-binding-domain vector as pointing from the centre of the first to the centre of the second.

\subsection{Potential energy}

Here, we reframe and refine the model as described in the original work~\cite{cumberworth2018_sup} in terms of a potential energy composed of three primary terms,
\begin{equation}
    U = U_{\textrm{bond}} + U_{\textrm{stack}} + U_{\textrm{steric}},
\end{equation}
where $U_{\textrm{bond}}$ is the contribution from bonding between binding domains, $U_{\textrm{stack}}$ is the contribution from base stacking between binding domains, and $U_{\textrm{steric}}$ is the contribution from steric interactions.
Both $U_{\textrm{stack}}$ and $U_{\textrm{steric}}$ are composed of further subterms that depend on two, three, or four bound-domain pairs.

\subsubsection{Bonding term}
\label{sec:bonding}

The energy of bound and misbound states is taken to be the unified-\ac{NN} model hybridization free energy~\cite{allawi1997_sup,santalucia1998_sup,santalucia2004_sup} of the two strands that occupy the lattice site.
We consider only fully hybridized segments, the relevant terms for which are
\begin{equation}
    \dgsnn  = \dgsi + \dgss + \sum \dgsst + \dgsat.
    \label{eq:nn}
\end{equation}
$\dgsi$ is a sequence-independent initiation free energy, $\dgsat$ is penalty for having a terminal AT pair, if applicable, $\dgss$ is a term that accounts for palindromic sequences (but since staples are designed never to be palindromic to prevent self binding, this term does not apply to bound domains), and $\dgsst$ is the stacking free energy, which is calculated for all (overlapping) pairs along the sequence.
The parameters of the unified-\ac{NN} model as given by \citet{santalucia2004_sup} assume a standard state amount concentration of \SI{1}{\Molar}.
Parameters are provided for both the enthalpic and entropic contribution, allowing for inclusion of the temperature dependence with $\dgsnn = \dhsnn - T\dssnn$, where $T$ is the temperature.
Sodium ion dependence can also be taken into account through an empirical relation,
\begin{equation}
    \dssnn \mleft( \mleft[ \textrm{Na}^+ \mright] \mright) = \dssnn + \frac{0.368N}{2} \ln \mleft( \frac{ \mleft[ \textrm{Na}^+ \mright]}{[\standardstate]} \mright),
    \label{eq:nn-salt}
\end{equation}
where $\dssnn$ is the standard-state entropy at \SI{1}{\Molar} NaCl, $N$ is the number of phosphate groups in the DNA strand, $\mleft[\textrm{Na}^+ \mright]$ is the sodium ion amount concentration, and $\mleft[ \standardstate \mright]$ is the standard state amount concentration.
Here, we assume the number of phosphates is equal to the number of nucleotides in the sequence.
In the case of partially complementary sequences, the hybridization free energy is approximated by the predicted free energy for the longest contiguous complementary sequence of the pair; this approximation has been shown to work well when simulating DNA bricks~\cite{wayment-steele2017_sup,reinhardt2014_sup,jacobs2015_sup}.

We refine the original model by formally mapping the hybridization free energies from the unified-\ac{NN} model to the interaction energy of two binding domains in our model; to do so, we follow the approach of \citet{reinhardt2016_sup}.
Here, we must consider two different types of reactions: intermolecular, in which a binding domain on a free staple hybridizes to a binding domain in a scaffold system, where a scaffold system refers to a single scaffold strand and any staples bound directly or indirectly to it, and intramolecular, in which two binding domains in a scaffold system hybridize.
In the first case, we have a bimolecular reaction with an equilibrium constant $K_\textrm{b}$ of the hybridization reaction between a staple-binding-domain A and a scaffold-system-binding-domain B,
\begin{gather}
    \label{eq:hybridization-reaction}
    \ce{A + B <=> AB},\\
    \label{eq:equilconstant}
    K_\textrm{b} = \frac{\ce{[AB][\standardstate]}}{\ce{[A][B]}} = \frac{C_\textrm{AB} C^\standardstate}{C_\textrm{A} C_\textrm{B}} = \textrm{e}^{-\beta \dgsnn},
\end{gather}
where $C_x$ is the number density (i.e.~number concentration) of $x$, $C^\standardstate = \textrm{N}_\textrm{A} \ce{[\standardstate]}$ is the standard state number density, $\textrm{N}_\textrm{A}$ is Avogadro's constant, and $\beta = 1 / \kb T$, with $\kb$ being the Boltzmann constant.
In equilibrium,
\begin{equation}
    \label{eq:equilcondition}
    \mu_\textrm{A} + \mu_\textrm{B} = \mu_\textrm{AB},
\end{equation}
where $\mu_x$ is the chemical potential of $x$.

To a first approximation, we can assume that the binding domains behave ideally with respect to each other, allowing the partition functions of the binding domains in our model, $Q_x \mleft(N_x, V, T \mright)$, to be expressed in terms of the internal partition functions of the binding domains, $q_x \mleft( V, T \mright)$,
\begin{equation}
    Q_x = \frac{V_\textrm{L}^{N_x}}{N_x!} q_x^{N_x},
\end{equation}
where $V_\textrm{L}$, the system lattice volume, is a dimensionless quantity as it represents a sum over all the lattice sites in the system, $N_x$ is the number of $x$ present in $V_\textrm{L}$, and the function notation on the partition functions has been dropped for notational simplicity.
Then using the relation
\begin{equation}
    \mu_x = -\kb T \mleft( \frac{\partial \ln Q_x}{\partial N_x} \mright)_{V,T}
\end{equation}
and Stirling's approximation, the chemical potential can be written in terms of the internal partition function,
\begin{equation}
    \mu_x = \kb T \ln \mleft( \frac{\rho_x}{q_x} \mright), \label{eq:muasqandrho}
\end{equation}
where $\rho_x$ is the lattice number density of $x$.
The lattice number density can be related to the number density with
\begin{equation}
    \rho_x = a^3 C_x, \label{eq:latticeconstant}
\end{equation}
where $a$ is the lattice constant, which has units of length.

Plugging in \cref{eq:muasqandrho} and \cref{eq:latticeconstant} to \cref{eq:equilcondition} and rearranging, we get
\begin{equation}
    \frac{C_\textrm{AB}}{C_\textrm{A} C_\textrm{B}} = a^3 \frac{q_\textrm{AB}}{q_\textrm{A} q_\textrm{B}}.
\end{equation}
Comparing to \cref{eq:equilconstant}, we can multiply through by $C^\standardstate$ to obtain
\begin{equation}
    \frac{q_\textrm{AB}}{q_\textrm{A} q_\textrm{B}} = \frac{\textrm{e}^{-\beta \dgsnn}}{a^3 C^\standardstate}. \label{eq:internalqratio}
\end{equation}
Given that each binding domain has an orientation vector with six possible configurations, the internal partition functions become
\begin{equation}
    q_\textrm{A} = q_\textrm{B} = 6, \quad q_\textrm{AB} = 6 \textrm{e}^{-\beta \varepsilon_\textrm{b}}, \label{eq:internalq}
\end{equation}
where $\varepsilon_\textrm{b}$ is the bimolecular binding domain interaction energy of our model.
Plugging in \cref{eq:internalq} to \cref{eq:internalqratio}, we can solve for $\varepsilon_\textrm{b}$,
\begin{gather}
    \frac{\textrm{e}^{-\beta \varepsilon_\textrm{b}}}{6} = \frac{\textrm{e}^{-\beta \dgsnn}}{a^3 C^\standardstate} \notag\\
    \rightarrow \varepsilon_\textrm{b} = \dgsnn + \kb T \ln \mleft( a^3 C^\standardstate \mright) - \kb T \ln 6.
    \label{eq:bi-epsilon}
\end{gather}

For the second case, where we are considering two binding domains already in the scaffold system, we have a unimolecular reaction with an equilibrium constant $K_\textrm{u}$ of the hybridization reaction between a system with two unbound domains C and a system with a bound-domain-pair D,
\begin{gather}
    \label{eq:uni-hybridization-reaction}
    \ce{C <=> D},\\
    \label{eq:uni-equilconstant}
    K_\textrm{b} = \frac{\ce{[C]}}{\ce{[D]}} = \frac{C_\textrm{C}}{C_\textrm{D}} = \frac{\rho_\textrm{C}}{\rho_\textrm{D}} = \textrm{e}^{-\beta \upDelta G^\standardstate_\textrm{NN, u}},
\end{gather}
where $\upDelta G^\standardstate_\textrm{NN, u}$ is the unimolecular unified-\ac{NN} standard Gibbs free energy of hybridization.
Because the $\dgsi$ term captures the translational entropy cost of combining two free strands into one~\cite{santalucia2004_sup}, we will assume that
\begin{equation}
    \upDelta G^\standardstate_\textrm{NN, u} = \dgsnn - \dgsi.
\end{equation}
We also assume that the scaffold systems act ideally with respect to each other, but now the treatment of the internal partition function is more complicated.
To make the problem tractable, we treat the binding domains within the scaffold system as being independent if they are not hybridized to each other.
Then, as before, we only need to consider the internal partition functions of the two unbound domains and the bound-domain pair, allowing us to solve for the unimolecular binding domain interaction energy $\varepsilon_\textrm{u}$,
\begin{gather}
    \frac{\rho_\textrm{D}}{\rho_\textrm{C}} = \frac{q_\textrm{D}}{q_\textrm{C}} = \frac{q_\textrm{AB}}{q_\textrm{A} q_\textrm{B}} = \frac{\textrm{e}^{-\beta \varepsilon_\textrm{u}}}{6} = \textrm{e}^{-\beta \upDelta G^\standardstate_\textrm{NN, u}} \notag\\
    \label{eq:uni-epsilon}
    \rightarrow \varepsilon_\textrm{u} = \upDelta G^\standardstate_\textrm{NN, u} - \kb T \ln 6.
\end{gather}
Because this is a unimolecular reaction, $\upDelta G^\standardstate_\textrm{NN, u}$ is not dependent on the standard state concentration, and so the model interaction energy does not need to have a term with the standard state concentration to be independent of changes to the standard state.

In order to calculate a chemical potential of a staple from a given concentration, we assume staples act ideally when in solution.
The canonical partition function for the staples of type $i$ is
\begin{equation}
    Q_i = \frac{ \mleft( q_i V_\textrm{L} \mright)^{N_i}}{N!} = \frac{ \mleft( 6^{2 n_i - 1} V_\textrm{L} \mright)^{N_i}}{N_i!},
\end{equation}
where $N_i$ is the number of staples of strand $i$, and $n_i$ is the number of binding domains that the staple strand comprises.
The chemical potential of staple strand $i$ is then
\begin{equation}
    \label{eq:chempot-ideal}
    \mu_i = \kb T \mleft[ \ln \mleft( a^3 C_i \mright) - \mleft(2 n_i - 1 \mright) \ln6 \mright].
\end{equation}

If we derive the melting temperature of the model for a single-binding-domain scaffold strand, we can compare it to the melting temperature of the unified-\ac{NN} model assuming a two state reaction to verify the derivation of our potential.
The average occupancy can be calculated exactly for this system, and it does not require the more complicated terms of the model detailed in the next sections.
We will calculate this value using the grand ensemble, where the system volume is the number of scaffold binding domains.
The grand partition function is
\begin{align}
    \Xi \mleft( \mu, V, T \mright) &= \sum_{N=0}^{1} \textrm{e}^{\beta \mu N} \sum_i \textrm{e}^{-\beta U_i} \notag\\
                                   &= \sum_i \mathrm{e}^{-\beta U_i} + \textrm{e}^{\beta \mu} \sum_i \mathrm{e}^{-\beta U_i} \notag\\
                                   &= 6 + \frac{6^2 \textrm{e}^{\beta \mu} \textrm{e}^{-\beta \dgsnn}}{a^3 C^\standardstate},
\end{align}
where the inner sum in the first line and the sums in the second line are over states with $N$ bound staples with potential energy $U_i$, and in the third line we have plugged in \cref{eq:bi-epsilon} and simplified.
The average occupancy is
\begin{align}
    \langle s \rangle &= \frac{1}{\Xi} \sum_N \textrm{e}^{\beta \mu N} \sum_i s \textrm{e}^{-\beta U_i} \label{eq:occupancy}\\
                      &= \frac{6 \textrm{e}^{\beta \mu} \textrm{e}^{-\beta \dgsnn}}{a^3 C^\standardstate + 6 \textrm{e}^{\beta \mu} \textrm{e}^{-\beta \dgsnn}},
\end{align}
where $s$ is 0 when the single scaffold binding domain is unbound and 1 when it is bound.
At the melting temperature, $T_\textrm{m}$, the average occupancy of the scaffold lattice site by a staple binding domain is $\nicefrac{1}{2}$, thus
\begin{align}
    1 &= \frac{\textrm{e}^{\beta \mu} 6 \textrm{e}^{-\beta \dgsnn}}{a^3 C^\standardstate} \notag\\
    1 &= \frac{C}{C^\standardstate} \textrm{e}^{-\beta \mleft( \dhsnn - T\dssnn \mright) } \notag\\
    \label{eq:model-melting-temp}
    \rightarrow T_\textrm{m} &= \frac{\dhsnn}{\kb \ln \mleft( \frac{C}{C^\standardstate} \mright) + \dssnn},
\end{align}
where in the second line we have used \cref{eq:chempot-ideal} with $n = 1$ and simplified.

The melting temperature for the unified-\ac{NN} model assuming a two state reaction can be derived directly.
If when considering \cref{eq:hybridization-reaction} we let A be the staple and B be the scaffold, and let $\ce{[A]}_\textrm{T} \geq \ce{[B]}_\textrm{T}$, where the subscript denotes the total concentration (i.e.~including the staple and scaffold binding domains when they are in the bound AB state), and if we consider an initial state with all B being bound in the AB form, then in equilibrium we have
\begin{equation}
    \label{eq:iceresults}
    \ce{[A]} = \ce{[A]}_\textrm{T} - \ce{[B]}_\textrm{T} + x, \quad \ce{[AB]} = \ce{[B]}_\textrm{T} - \ce{[B]}.
\end{equation}
At the melting temperature, $\ce{[B]} = \ce{[AB]}$; plugging in this and \cref{eq:iceresults} to \cref{eq:equilconstant} and rearranging gives
\begin{align}
    K &= \frac{\ce{[A]}_\textrm{T}}{[\standardstate]} - \frac{\ce{[B]}_\textrm{t}}{2[\standardstate]} = \textrm{e}^{-\beta \dgsnn}\\
    \rightarrow T_\textrm{m} &= \frac{\dhsnn}{\kb \ln \mleft( \frac{\ce{[A]}_\textrm{T}}{[\standardstate]} - \frac{\ce{[B]}_\textrm{T}}{2[\standardstate]} \mright) + \dssnn} \notag\\
                             &\simeq \frac{\dhsnn}{\kb \ln \mleft( \frac{C_\textrm{A}}{C^\standardstate} \mright) + \dssnn},
\end{align}
where the asymptotic equality follows when $\ce{[A]}_\textrm{T} \gg \ce{[B]}_\textrm{T}$, which we assume in our model.
Comparing this with \cref{eq:model-melting-temp}, we see that the melting temperatures agree.

While the model melting temperature agrees with the unified-\ac{NN} two-state melting temperature for a single-binding-domain scaffold, it will not hold for anything longer because of the oversimplified assumption of the internal partition function of the scaffold-system binding domains being independent.
The internal partition function of the system is highly non-trivial, so the best we can do is use a mean field approach to give an average difference of the logarithms of the partition functions with a change in the binding state of the system.
It is important to keep in mind, however, that even if we could calculate the ratio of the partition functions exactly in order to correct the unified-\ac{NN} hybridization free energy for every possible hybridization reaction, we would be creating a model in which the individual binding domains hybridize with the same statistics as the unified-\ac{NN} model.
This is not the expected behaviour for DNA-origami binding domains.
It is precisely the deviation from the \ac{NN} model in these internal hybridization reactions that we are interested in studying.
It is here that the advantage of using a model with a physical basis over more statistical models becomes apparent, as these deviations, which are entropic in nature, are naturally present up to some constant, and so the entropy differences will be roughly captured.

To understand why some correction is still needed, consider that while the cooperativity involved in DNA origami self-assembly is expected to change the overall slope of a curve of an order parameter as a function of the temperature, the curve should not be shifted to overall higher or lower melting temperatures relative to a pure unified-\ac{NN}.
If a fully assembled state has only one allowed configuration, then without further modification, the model as defined will have a melting temperature that is dependent on the choice of binding-domain size.
Consider a particular design represented in two different ways, where the second has binding domains defined as being twice as small as the first.
Because in both cases the assembled state has just one configuration, the loss in entropy will be twice as large for the second system, which will shift the assembly to lower temperatures.

The above mentioned mean field correction can allow for such an overall correction.
However, fully assembled states will not in general have only one configuration in our model.
The number of states available will depend on how the fully assembled state is defined, how the remaining terms of the potential are defined, and on the specific design being considered.
We can begin by considering the most extreme case, where there is only one configuration available in the bound state to give an upper bound on the absolute value of the correction.
If each binding domain has six relative positions and six orientation vectors, then upon hybridization of two binding domains,
\begin{gather}
    q_\textrm{A} = q_\textrm{B} = 6^2, \quad q_\textrm{AB} = \textrm{e}^{-\beta \varepsilon_\textrm{u}} \notag\\
    \frac{\rho_\textrm{D}}{\rho_\textrm{C}} = \frac{q_\textrm{D}}{q_\textrm{C}} = \frac{q_\textrm{AB}}{q_\textrm{A} q_\textrm{B}} = \frac{\textrm{e}^{-\beta \varepsilon_\textrm{u}}}{6^4} = \textrm{e}^{-\beta \upDelta G^\standardstate_\textrm{NN, u}} \notag\\
    \rightarrow \varepsilon_\textrm{u} = \upDelta G^\standardstate_\textrm{NN, u} - 4 \kb T \ln 6.
\end{gather}
For binding of a staple to a partially assembled scaffold, we will have a different expression, as the first binding event is a change in absolute position rather than relative position,
\begin{gather}
    q_\textrm{A} = 6, \quad q_\textrm{B} = 6^2, \quad q_\textrm{AB} = \textrm{e}^{-\beta \varepsilon_\textrm{b}} \notag\\
    \frac{q_\textrm{AB}}{q_\textrm{A} q_\textrm{B}} = \frac{\textrm{e}^{-\beta \varepsilon_\textrm{b}}}{6^3} = \frac{\textrm{e}^{-\beta \dgsnn}}{a^3 C^\standardstate} \notag\\
    \rightarrow \varepsilon_\textrm{b} = \dgsnn + \kb T \ln \mleft( a^3 C^\standardstate \mright) - 3 \kb T \ln 6.
\end{gather}
Finally, special consideration must be made for the overall system's rotational entropy, which is not lost in the final assembled state.
If one considers the first three binding domains of the scaffold, the second will always have six relative positions to the first by rotation of the entire system, and the third will always have four positions relative to the second by rotation of the entire system around the bond axis between the first and the second binding domains.
There is also no relative positional entropy to lose when binding the first scaffold binding domain.
Thus, for the first staple to bind to the scaffold, we add $2 \kb T \ln6$ to $\varepsilon_\textrm{b}$, while for the second scaffold domain to bind, whether to another binding domain on the first staple or to a new staple, we add $\kb T \ln6 - \kb T \ln2 = \kb T \ln 3$ to either $\varepsilon_\textrm{u}$ or $\varepsilon_\textrm{b}$, respectively.

\subsubsection{Stacking term}
\label{sec:stacking}
\begin{figure*}[t!]
    \centering
    \includegraphics{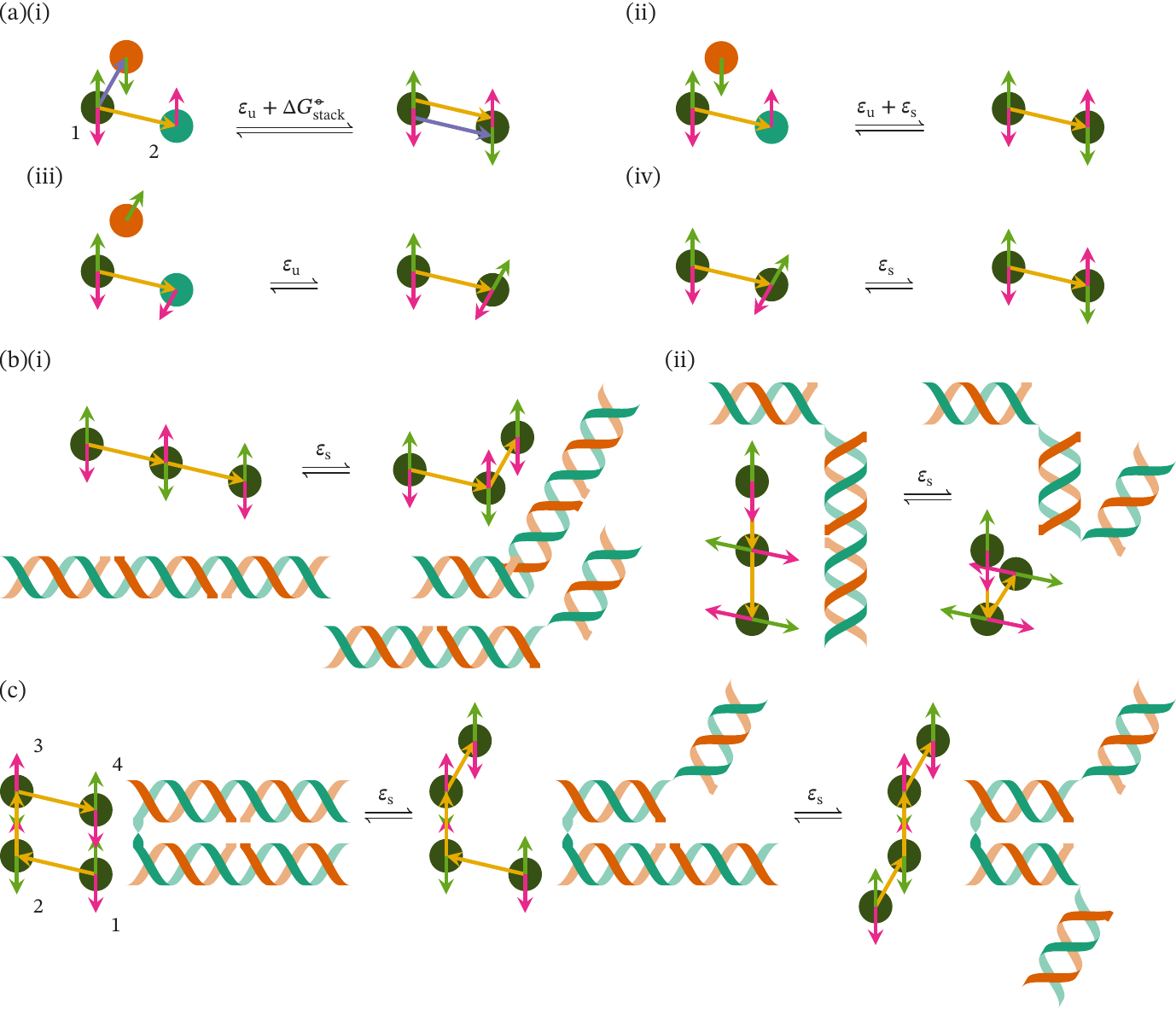}
    \caption{
        \label{fig:stacking}
        Helical stacking in the model.
        (a) Helical stacking with two bound-domain pairs.
        In (i), there are two pairs of contiguous domains binding to each other, such that a single helix is formed.
        In (ii)--(iv), only one pair of binding domains is contiguous, allowing for unstacked, or kinked, configurations to form.
        While not drawn, we assume here that the staple binding domain is part of a staple that is bound by another one of its domains to the scaffold system.
        While stacking and bonding can occur in one step as in (ii), it is also possible for bonding to occur first, as in (iii), followed by stacking (iv).
        (b) Helical stacking with three bound-domain pairs.
        In (i), two model configurations are shown that are both pairwise stacked, but the configuration on the right has one less stacking interaction.
        In (ii), two model configurations are shown with just one pairwise stack, but the configuration on the right has one less stacking interaction.
        (c) Helical stacking with four bound-domain pairs.
        All three configurations have pairwise stacks, but the configuration in the middle has only one stacking interaction, and the one on the right has none.
    }
\end{figure*}

If two contiguous domains are in bound states and part of the same helix, then there is an additional stacking term that we have not yet accounted for when calculating the unified-\ac{NN} model hybridization free energy for the binding domains separately.
There are two cases to consider at the level of resolution of our model.
The first case is that where there are only two strands involved, with each having two contiguous binding domains, in which case, as discussed in \cref{sec:bonding}, there is only one allowed configuration for the orientation vectors involved (\cref{fig:stacking}(a)(i)).
Here, it is reasonable to add the $\dgsst$ corresponding to the nucleotides involved to \cref{eq:uni-epsilon}, and the mean field entropy correction discussed above would now be a local description for the binding of the second pair of domains.

The second case is that where only one pair of binding domains is contiguous (\cref{fig:stacking}(a)(ii)--(iv)).
This could involve one strand ending and another beginning (this is sometimes referred to as a nick in the backbone of one of the strands forming the helix), or one or both may continue and possibly even be a part of another helix via a crossover junction.
We will refer to the point at which any of these situations occur as breakpoints.
Breakpoints are able to become unstacked to form kinks.
In the model, if the orientation vectors of a pair of contiguous bound domains do not have a configuration prescribed by the helical geometry, they are considered to have a kink and are treated as two separate helices (for an example see \cref{fig:helical-twist}(a)(ii)).
In other words, the pair of bound domains can either be in a stacked configuration or a kinked configuration, and the definition that was given for when a pair of bound domains can be identified as being part of the same helix can also be be used to identify stacked states.
In contrast to the first case in which there is no breakpoint, the stacking and domain binding events are independent, and a separate stacking energy term is required, $\varepsilon_\textrm{s}$ (\cref{fig:stacking}(a)(iii) and (iv)).
As this is actually a free energy describing the difference between being stacked and kinked, it will, in principle, be different from the $\dgsst$ term.

Whether or not two pairs of bound domains are stacked or kinked cannot always be determined by considering pairwise interactions in our model.
This is because the helical axis is defined implicitly, as discussed in \cref{sec:state-space}.
Consider three bound-domain pairs that occupy adjacent lattice sites, in which a strand is contiguous between both pairs of adjacent lattice sites (but not necessarily the same strand across all three).
There are multiple configurations for which the pairwise stacking rule is obeyed for both pairs of bound-domain pairs, two of which are shown in (\cref{fig:stacking}(b)(i)).
However, all but one of these configurations involve a right-angle bend in the helix.
The length of the binding domains is typically well below the persistence length of double-stranded DNA, so configurations with such sharp turns are extremely unlikely with no external force.
Therefore, these configurations must have a kink at one of the breakpoints, and are in fact composed of two separate helices.
Such configurations will then only have one of the two stacking interaction energies.
If there are two breakpoints, there is ambiguity in which breakpoint is actually kinked, and so the stacking interaction in our model is not entirely local (\cref{fig:stacking}(b)(i)).

If we consider the definition of the next-binding-domain vector as discussed above, we note that there are configurations for two bound-domain pairs with orthogonal helices that also map to model configurations defined as being stacked.
Configurations only map to one model configuration, so in effect these configurations are not included in the model.
We have made this choice to allow stacking to be defined between pairs of bound domains without adding additional degrees of freedom to the model.
Once there are more than two bound-domain pairs involved, these configurations are included by applying the stacking term described above and shown in \cref{fig:stacking}(b)(i).
However, the model has no way to represent both of these kinks occurring at the same time.
Considering that there are many ways for the model to represent kinked configurations, this exclusion seems reasonable to allow for a simpler model.

For configurations in which the second binding-domain's helix is orthogonal to the first binding-domain's helix, it is not possible for a third binding domain to form a stacked configuration with the second binding domain and for that resulting helix to be in the same plane as the first.
However, without any further terms, it is possible to construct model configurations in which this is the case; an example is given in \cref{fig:stacking}(b)(ii).
In order to make the model consistent, an additional term could be applied such that these configurations, while containing a pairwise stacked configuration, would be defined to have no stacking interaction.
However, for quarter- and three-quarter-turn binding domains, one of the model configurations between two bound-domain pairs can be mapped to from either a helix that is orthogonal or parallel to the first binding-domain's helix (see below for further discussion of all sterically allowed configurations and their mappings).
Because the configurations involving stacked parallel helices with crossovers are critical to origami designs, it is important not to apply a term that prevents stacking of additional binding domains.
A simple solution is to not remove the pairwise stacking interaction in configurations in which this ambiguity exists; these are configurations in which the first binding-domain's helical axis is equal to its orientation vector (again see below for further discussion).

For configurations in which the second binding-domain's helix is parallel to the first (i.e.~those which are involved in crossovers, which are important to represent correctly to model assembled configurations), there is additional complexity in determining whether configurations are stacked or kinked.
These configurations map to model configurations in which the first binding domain's next-binding-domain vector $\uvec{n}_1$ is equal to its orientation vector $\uvec{o}_1$, $\uvec{n}_1 = \uvec{o}_1$.
When both bound-domain pairs on either side of the breakpoint have another bound-domain pair that is contiguous to at least one of the involved strands, it is possible to construct model configurations which have two pairwise stacks that map to configurations that have only one or no stacked bound-domain pairs.
Using the indices in \cref{fig:stacking}(c), if $\uvec{n}_1 = -\uvec{n}_3$, then there are two stacking interactions.
If $\uvec{n}_1 \perp \uvec{n}_3$, then there is one stacking interaction.
If $\uvec{n}_1 = \uvec{n}_3$, then there are no stacking interactions.

\subsubsection{Steric term}

\begin{figure*}[t!]
    \includegraphics{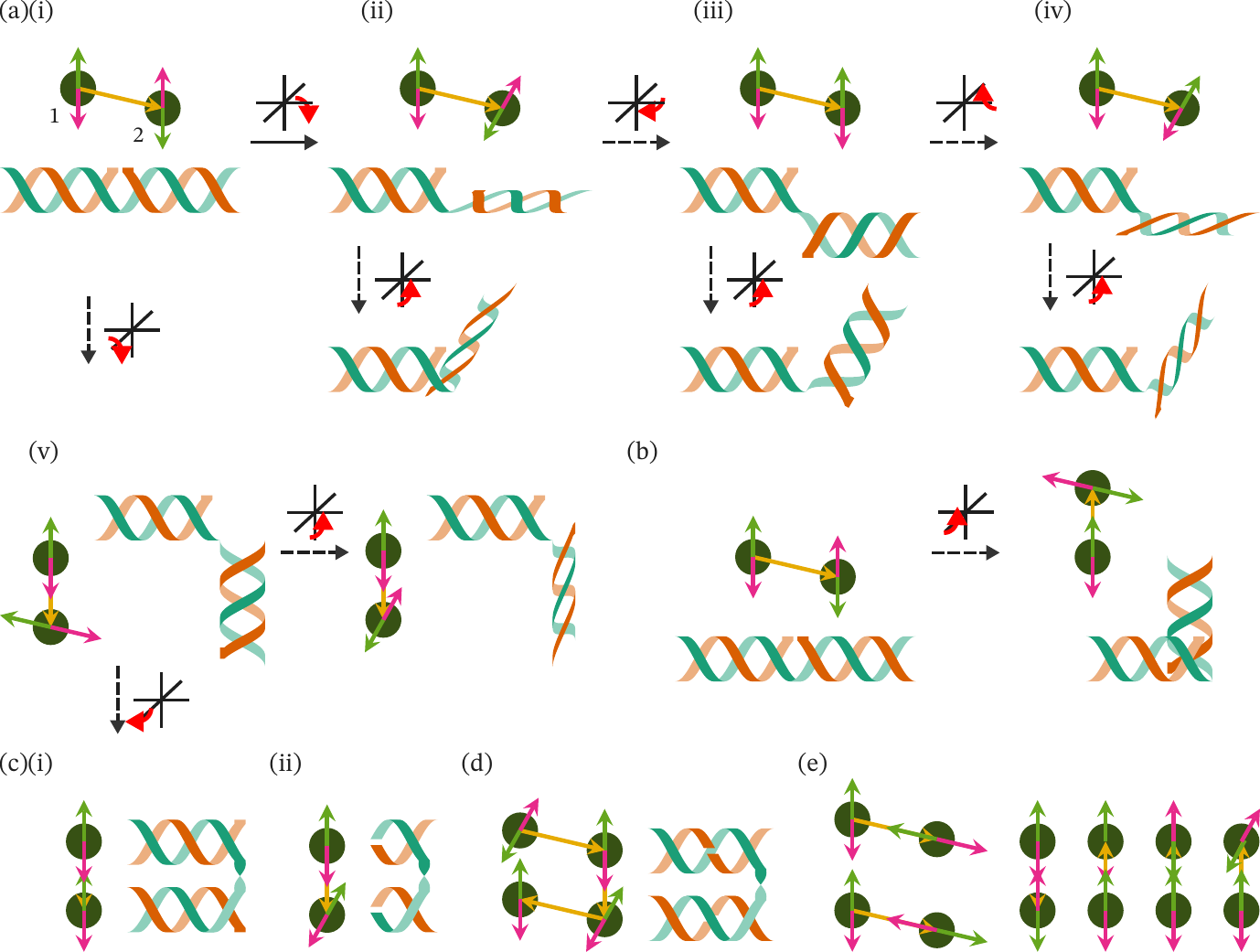}
    \caption{
        \label{fig:steric-pairs}
        All twelve unique configurations of two bound-domain pairs with a breakpoint.
        (a) Five configurations that are sterically allowed for all binding-domain classes.
        Cartoon helix representations are shown for a 16-\ac{nt} half-turn binding domain.
        (b) An example of a transformation that results in a sterically prohibited configuration.
        (c) A half-turn around the third rotation axis produces an additional allowed model configuration for half-turn binding domains (i), but not for quarter- or three-quarter-turn binding domains (ii).
        In (ii), a cartoon helix representation for an 8-\ac{nt} three-quarter-turn binding domain is shown.
        (d) Sterically allowed configuration for an 8-\ac{nt} three-quarter-turn binding domain with two pairs of stacked bound-domain pairs connected via a crossover.
        (e) All six sterically prohibited configurations.
    }
\end{figure*}

In describing the steric terms, we refer to `constraints' and `rules', but it should be understood that formally we are defining configurations that obey these constraints or rules as having a potential energy of zero for the term in question, and all others as having a potential energy of infinity.
In principle these could also form a part of the definition of state space, as was done for the rule that orientation vectors on bound domains must be opposing, but we have found it more convenient to define these as part of the potential.

Because it is directly related to the discussion of stacked configurations above, we again consider three bound-domain pairs on adjacent lattice sites.
In particular, we consider the case in which there are no breakpoints, which occurs when there are two triplets of binding domains that are contiguous on the same strand.
Again, there are multiple configurations for which the pairwise stacking rule is obeyed for both pairs of bound-domain pairs.
Unlike the case when there is at least one breakpoint, it is not possible for there to be a kink to allow for the configurations that have a right angle bend.
Therefore, all configurations but that in which there is no bend in the helix are disallowed.

Consider a pair of contiguous bound domains with a breakpoint.
In reality, the breakpoint will not allow for all possible relative orientations of the two binding domains.
By considering transformations to the two helical binding domains and making simple steric arguments, we can introduce further rules to account for this.
Because the model is already so coarse, the particular choices made are unlikely to have much effect beyond changing the entropic balance between bound and unbound states, unless they affect whether crossovers are only able to occur where they are allowed.
Thus it is sufficient to base our steric arguments on considerations of idealized cartoon helices.
The correct entropic balance could be restored by considering a correction factor to the free energies of hybridization that could be determined by comparison to experiment or simulations with a finer resolution, although we do not do so in this work.

Another consideration in constructing these terms is that by using more constrained potentials, sampling can become more difficult because the free-energy landscape becomes more rough.
For example, in the extreme case of not allowing kinked configurations, which was a form the model took in the early stages of development, sampling was very difficult as typically to rearrange the structure, the domains had to unbind and rebind.
Therefore, the guiding principle in constructing the potential for kinked configurations was to ensure that crossovers between parallel helices only occur at the correct intervals, to make the partially assembled structures as unconstrained as possible, and to achieve what physical realism we can with the steric arguments.

For all unique configurations involving two bound-domain pairs that have a breakpoint between them, all sterically allowed model configurations are illustrated in \cref{fig:steric-pairs}(a) and (c) and all pairwise sterically prohibited configurations are illustrated in \cref{fig:steric-pairs}(e).
For helix cartoon configurations drawn in \cref{fig:steric-pairs}, a 16-\ac{nt} half-turn binding domain is used, but the general arguments here hold for all three binding-domain classes.
To understand which configurations are possible, we must consider a number of rotations of the second binding-domain's helix relative to the first binding-domain's helix.
Beginning from a stacked configuration, consider rotating the second binding-domain's helix around an axis parallel to the helical axis but displaced to the outside of the helix to produce the configurations in \cref{fig:steric-pairs}(a)(ii)--(iv).
We will refer to this as the first rotation axis.
This allows for configurations in which $\uvec{n}_1 \perp (\uvec{o}_1 \land \uvec{o}_2$).

Following this first rotation with further rotations of the second binding-domain's helix around an axis parallel to the orientation vector of the first, which we will refer to as the second rotation axis, will not lead to any new relative orientations of $\uvec{o}_2$ because of our definition of mapping binding domains that form orthogonal helices to lattice sites.
An example of the resulting cartoon helix configuration after rotating in one direction is shown below the first cartoon helix configuration in \cref{fig:steric-pairs}(a)(ii)--(iv).
There are no cartoon helix configurations that map to model configurations in which $\uvec{o}_2 = \pm \uvec{n}_1$.
Thus, in our model, if $\uvec{n}_1 \perp \uvec{o}_1$, then $\uvec{o}_2 \perp \uvec{n}_1$.

\begin{figure*}[t!]
    \centering
    \includegraphics{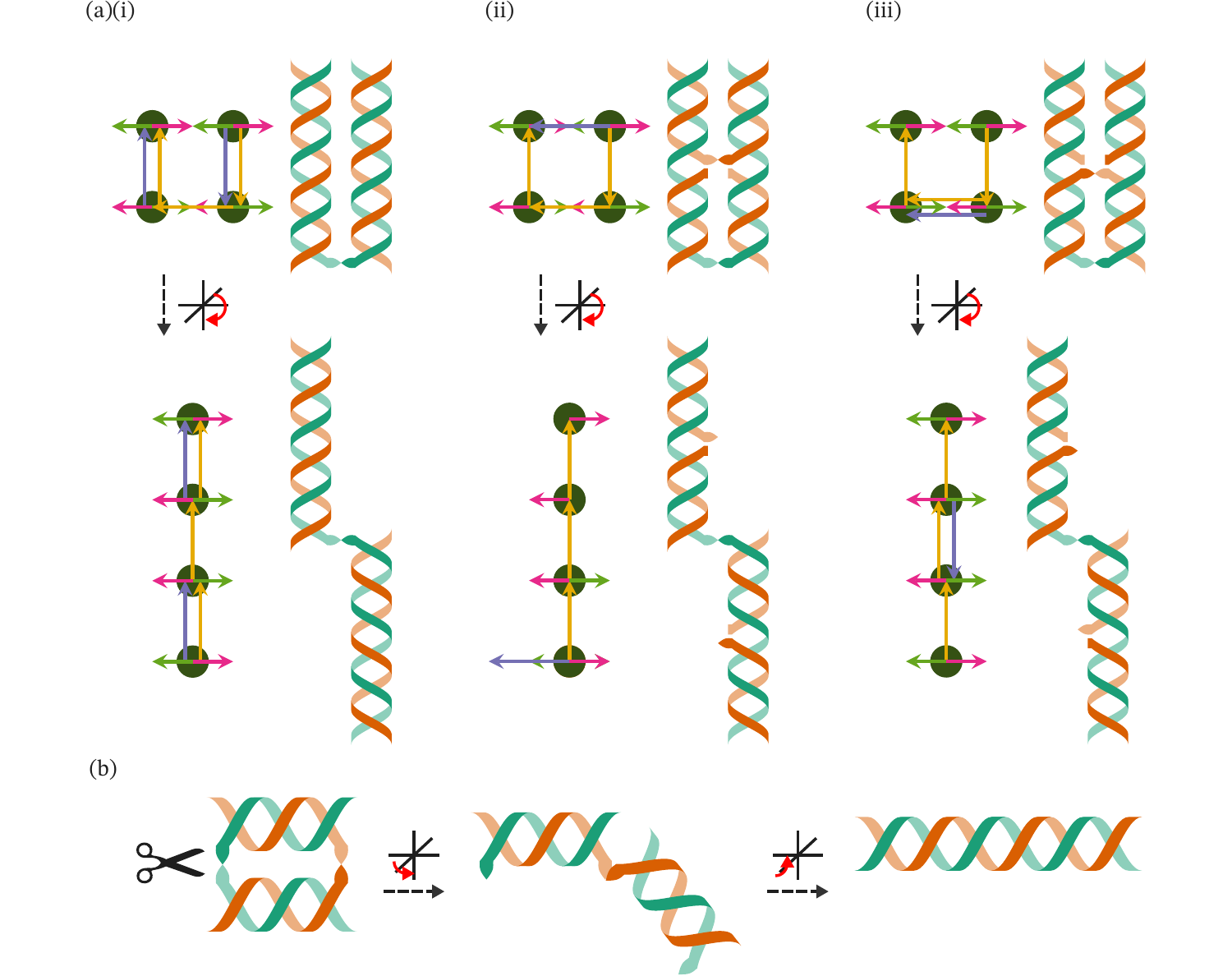}
    \caption{
        \label{fig:crossovers}
        Strand crossovers between helices involving 16-\ac{nt} binding domains.
        (a) Crossovers between two adjacent parallel helices with a four-binding-domain scaffold.
        (i) Helices with a single crossover.
        (ii) Helices with two crossovers on separate binding domains.
        (iii) Helices with crossovers on the same binding domain.
        (b) Doubly contiguous domains in bound states.}
\end{figure*}

Rotations around an axis perpendicular to the two previously mentioned rotation axes, which we will refer to this as the third rotation axis, can lead to configurations in which $\uvec{n}_1 = \pm \uvec{o}_1$.
If a quarter turn is made such that $\uvec{n}_1 = -\uvec{o}_1$, it leads to configurations with steric clashes, which is illustrated in \cref{fig:steric-pairs}(b).
A further quarter turn will only lead to worsening the steric clashes.
Thus, configurations in which $\uvec{n}_1 = -\uvec{o}_1$ are entirely disallowed.
Returning to the original stacked configuration and making a quarter turn around the third rotation axis in the opposite direction will lead to a configuration in which the second binding-domain's helical axis is orthogonal to the first \cref{fig:steric-pairs}(a)(v).
Unlike the previous orthogonal helix configurations, this configuration maps to a new allowed model configuration.
Rotations of this configuration around the second rotation axis will generate configurations that map to the same model configuration.

A further quarter turn around the third rotation axis leads to configurations in which the second binding-domain's helical axis is parallel to the first binding-domain's helical axis \cref{fig:steric-pairs}(c).
Such configurations are those that allow for crossovers between parallel helices, which are prevalent in the assembled state.
$\uvec{o}_2$ will depend on the length of the binding domain in these configurations.
In general, relative to $\uvec{o}_1$, $\uvec{o}_2$ will form the dihedral angle prescribed by the number of base pairs per bound-domain pair along the first helical axis, followed by a flip in the plane normal to $\uvec{o}_1$.
In the case of the half-turn binding domains, $\uvec{o}_2 = \uvec{o}_1$, producing a unique model configuration.

For quarter- and three-quarter-turn binding domains this will result in configurations that map to the model configuration in \cref{fig:steric-pairs}(a)(v), respectively, so no new model configurations are produced, and neither configurations where $\uvec{o}_2 = \pm \uvec{n}_1$ are sterically allowed.
That the crossover model configuration has more than one cartoon helix configuration that maps to it means there is a loss of information about the helical phase.
Both the second cartoon helix of \cref{fig:steric-pairs}(a)(v) and a cartoon helix configuration in which the second binding domain is rotated a half turn around the second rotation axis map to this model configuration, but only one correctly describes the crossover configuration.
To deal with this will introduce an additional term that applies to configurations in which both bound-domain pairs are stacked with an adjacent bound-domain pair.
Then, all four orientation vectors must be in the same configuration as they would be if all four bound-domain pairs were stacked in a single helix (\cref{fig:steric-pairs}(d)).
This term is only critical if crossovers occur such that the final structure is not planar, as otherwise the information loss on the phase has no effect on the assembled structures.

When there is more than one crossover between two helices, the helices become much more restricted in the configurations they are able to take relative to each other (compare \cref{fig:crossovers}(a)(i) to (ii)).
In particular, they will be forced to be roughly parallel.
This is naturally captured by the model when there are crossovers between more than one set of bound-domain pairs on two separate helices, as seen in \cref{fig:crossovers}(b)(ii).
However, when a single bound-domain pair has a double crossover, something which can occur with  half-turn binding domains, this will not be captured by the model as currently defined.

Consider two adjacent lattice sites in bound states, where at least one pair of binding domains are contiguous.
If the other pair of binding domains are not contiguous and in the same helix, their orientation vectors will still satisfy the prescribed helical angle because of the requirement of their orientation vectors to be opposing those of the strand that has two contiguous binding domains in that helix.
However, the case in which both pairs of binding domains are contiguous requires further consideration.
In reality, if the combined sequence of the two binding domains on one chain is together the reverse complement of the combined sequence of the two binding domains on the other chain, then the only way for all binding domains to be bound to each other is if there is only one helix.
If instead the binding domains on one chain must be swapped to make the whole two-binding-domain sequence the reverse complement of the other whole two-binding-domain sequence, then the only way for all binding domains to be bound to each other is if there are two parallel helices with both strands crossing over.
As a concrete illustration, one of the chains would have to be cut and glued to its other end to transition between these two configurations (\cref{fig:crossovers}(b)).
Thus, the model constrains pairs of contiguous complementary binding domains bound to each other to be in the same helix if they are the full reverse complements of each, and to be crossing over if not.

\section{Simulation and analysis methods}

\subsection{Ensemble and move types}

As in the study that first introduced the DNA-origami lattice model~\cite{cumberworth2018_sup}, we assume the staples do not interact when free in solution and are present in excess of scaffold strands such that their concentration can be assumed to be constant.
We also use the grand ($\mu V T$) ensemble to improve simulation efficiency.
A single scaffold is present in the simulation, so interactions between different scaffolds are not considered.
We also use the move types that were developed in that same work.
Specifically, we use an orientation rotation move type, a staple exchange move type, a \ac{CB} staple regrowth move type, and both a contiguous and non-contiguous \ac{CTRG} scaffold regrowth move type.
The associated move type parameters were set to the same as those in used in Reference \cite{cumberworth2018_sup}.

\subsection{Order parameters}

\begin{figure*}[t]
    \includegraphics{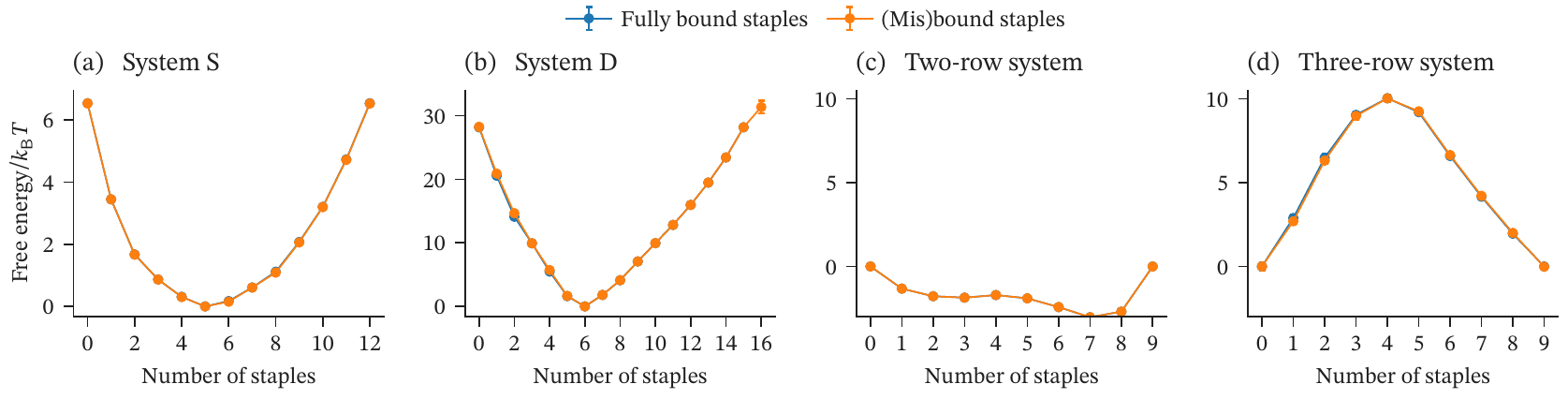}
    \caption{
        \label{fig:numstaples}
        Free energies for the number of fully bound staples and the number of (mis)bound staples.
        The two order parameters give nearly identical free energy profiles.
    }
\end{figure*}

To quantify the progress of assembly, we must define order parameters that allow us to construct a free-energy landscape.
There are several possible order parameters that could be defined.
Here, we consider three: the number of (partially or fully) bound staples, the number of fully bound staples, and the number of bound-domain pairs.
The first order parameter counts the total number of staples bound in some way to the system, whether bound to a fully complementary domain on the scaffold, or misbound to the scaffold or to another staple in the system; we will refer to this as the number of (mis)bound staples.
This order parameter is perhaps the closest analogue of that used in simulation studies of DNA bricks, i.e.~the number of bricks in the largest cluster~\cite{reinhardt2014_sup,reinhardt2016_sup,reinhardt2016b_sup,wayment-steele2017_sup}.
However, one possible problem with this order parameter is that it cannot be used to determine whether the system is fully assembled, as it does not measure the extent to which the scaffold is correctly folded.
For example, even for a completely unfolded scaffold, it is possible that each staple is bound to only one of its binding domains.
Such a configuration is clearly not at all folded, yet it cannot be differentiated from an assembled state with the same order parameter.
A related order parameter which circumvents this issue is the number of fully bound staples, where a fully bound staple is one in which all of its binding domains are bound to the correct scaffold binding domains.
In practice, however, when we compute free energies, we see that the sort of aberrant states that one can envisage do not appear to matter, and the two order parameters result in essentially identical free energies for all systems considered, indicating that staples bind either fully or not at all (\cref{fig:numstaples}).

To achieve a higher-resolution view of the binding of each staple, we can use the total number of bound-domain pairs as an order parameter.
However, this order parameter also cannot be used in isolation to determine if the system is in an assembled state, as multiple staples of a given type may bind to give the same number of bound-domain pairs as in the assembled state, which is known as a blocked state~\cite{snodin2016_sup}.

\subsection{Replica exchange}

\Ac{REMC}~\cite{swendsen1986_sup,geyer1991_sup,tesi1996_sup,hukushima1996_sup,yan1999_sup,sugita2000_sup,fukunishi2002_sup,earl2005_sup} involves running multiple replicas in parallel, differing by one or more control variables (e.g.~temperature).
The replica exchange step attempts to swap the configurations between a pair of replicas, typically those that are adjacent with respect to the control variables.
Instead of attempting to exchange at a set step interval, the scheme here alternates between attempting an exchange between all even pairs and all odd pairs of replicas, where pairs are numbered with the index of the first replica in the pair along the control variable~\cite{manousiouthakis1999_sup,lingenheil2009_sup}.
In the \ac{REMC} simulations performed in this study, because of the temperature dependence of the hybridization free energies, temperature \ac{REMC} also involves a change in the Hamiltonian.
Further, because we are in the grand ensemble but would like to keep the staple concentration constant across the replicas, we must also consider the change in staple chemical potential.

Considering each replica as its own simulation, a \ac{REMC} swap move can be considered as two separate moves for the selected replica pair, $i$ and $j$.
For replica $i$ the detailed balance condition is
\begin{multline}
    \label{eq:remc-detailed-balance-i}
    p \mleft( \vec{y}; \mu_i, T_i, \mathcal{H}_i \mid \vec{x}; \mu_i, T_i, \mathcal{H}_i \mright) p \mleft( \vec{x}; \mu_i, T_i, \mathcal{H}_i \mright)\\
    = p \mleft( \vec{x}; \mu_i, T_i, \mathcal{H}_i \mid \vec{y}; \mu_i, T_i, \mathcal{H}_i \mright) p \mleft( \vec{y}; \mu_i, T_i, \mathcal{H}_i \mright),
\end{multline}
while for replica $j$ it is
\begin{multline}
    \label{eq:remc-detailed-balance-j}
    p \mleft( \vec{x}; \mu_j, T_j, \mathcal{H}_j \mid \vec{y}; \mu_j, T_j, \mathcal{H}_j \mright) p \mleft( \vec{y}; \mu_j, T_j, \mathcal{H}_j \mright)\\
    = p \mleft( \vec{y}; \mu_j, T_j, \mathcal{H}_j \mid \vec{x}; \mu_j, T_j, \mathcal{H}_j \mright) p \mleft( \vec{x}; \mu_j, T_j, \mathcal{H}_j \mright).
\end{multline}
For the swap to be accepted, both of these individual moves must occur, so the total transition probability is the product of the two individual transition probabilities, giving the detailed balance condition
\begin{widetext}
\begin{multline}
    \label{eq:remc-transitions}
    p \mleft( \vec{y}; \mu_i, T_i, \mathcal{H}_i \land \vec{x}; \mu_j, T_j, \mathcal{H}_j \mid \vec{x}; \mu_i, T_i, \mathcal{H}_i \land \vec{y}; \mu_j, T_j, \mathcal{H}_j \mright)\\
    = p \mleft( \vec{y}; \mu_i, T_i, \mathcal{H}_i \mid \vec{x}; \mu_i, T_i, \mathcal{H}_i \mright) p \mleft( \vec{x}; \mu_j, T_j, \mathcal{H}_j \mid \vec{y}; \mu_j, T_j, \mathcal{H}_j \mright)
\end{multline}
\end{widetext}
We can then rewrite the transition probability in terms of the acceptance and trial probabilities.
The generation of a trial configuration for a replica is essentially taking a configuration from the equilibrium ensemble for its selected pair for that move, so
\begin{equation}
    \label{eq:remc-trial}
    p_{\textrm{trial}} \mleft( \vec{y}; \mu_i, T_i, \mathcal{H}_i \mid \vec{x}; \mu_i, T_i, \mathcal{H}_i \mright) = p \mleft( \vec{y}; \mu_j, T_j, \mathcal{H}_j \mright).
\end{equation}
Combining \cref{eq:remc-detailed-balance-i,eq:remc-detailed-balance-j,eq:remc-transitions,eq:remc-trial} and rearranging gives
\begin{widetext}
\begin{align}
    \label{eq:remc-acc}
    p_\textrm{acc} \mleft( \vec{y}; \mu_i, T_i, \mathcal{H}_i \land \vec{x}; \mu_j, T_j, \mathcal{H}_j \mid \vec{x}; \mu_i, T_i, \mathcal{H}_i \land \vec{y}; \mu_j, T_j, \mathcal{H}_j \mright) &= \min \mleft[ 1,\ \frac{p \mleft( \vec{y}; \mu_i, T_i, \mathcal{H}_i \mright) p \mleft( \vec{x}; \mu_j, T_j, \mathcal{H}_j \mright)}{p \mleft( \vec{y}; \mu_j, T_j, \mathcal{H}_j \mright) p \mleft( \vec{x}; \mu_i, T_i, \mathcal{H}_i \mright)} \mright] \notag \\
    &= \min \mleft[ 1,\ \frac{ \left( \textrm{e}^{\beta_i \mu_i N_{\vec{y}}}\textrm{e}^{-\beta_i \mathcal{H}_i \mleft( \vec{y} \mright)} \right) \left( \textrm{e}^{\beta_j \mu_j N_{\vec{x}}}\textrm{e}^{-\beta_j \mathcal{H}_j \mleft( \vec{x} \mright)} \right) }{ \left( \textrm{e}^{\beta_j \mu_j N_{\vec{y}}}\textrm{e}^{-\beta_j \mathcal{H}_j \mleft( \vec{y} \mright)} \right) \left( \textrm{e}^{\beta_i \mu_i N_{\vec{x}}}\textrm{e}^{-\beta_i \mathcal{H}_i \mleft( \vec{x} \mright)} \right) } \mright] \notag \\
    &= \min \mleft[ 1,\ \textrm{e}^{\upDelta_{\textrm{r}} \mleft( \beta \upDelta_{\textrm{c}} \mathcal{H} \mright) - \upDelta_{\textrm{r}} \mleft( \beta \mu \mright) \upDelta_{\textrm{c}} N} \mright]
\end{align}
\end{widetext}
for the acceptance probability, where $\upDelta_{\textrm{r}}$ is a difference operator between replicas $i$ and $j$ (e.g.~$\upDelta_{\textrm{r}} \mleft( ab \mright) = a_j b_j - a_i b_i$) and $\upDelta_{\textrm{c}}$ is a difference operator between configurations $\vec{x}$ and $\vec{y}$ (e.g.~$\upDelta_\textrm{c} a = a_{\vec{y}} - a_{\vec{x}}$).

From \cref{eq:remc-acc}, it can be seen that to calculate the acceptance probability for a given swap, it is necessary to calculate the energy of both configurations with both Hamiltonians.
If we expand the Hamiltonian in the first term of the exponential in \cref{eq:remc-acc} in terms of the enthalpy and entropy of the model, we can simplify this calculation such that
\begin{widetext}
\begin{align}
    \upDelta_{\textrm{r}} \mleft( \beta \upDelta_{\textrm{c}} \mathcal{H} \mright) & = \frac{1}{\kb T_j} \left( \upDelta H_{\textrm{total}} \mleft( \vec{y} \mright) - T_j \upDelta S_{\textrm{hyb}} \mleft( \vec{y} \mright) - \upDelta H_{\textrm{total}} \mleft( \vec{x} \mright) + T_j \upDelta S_{\textrm{hyb}} \mleft( \vec{x} \mright) \right) \notag \\
    & \quad - \frac{1}{\kb T_i} \left( \upDelta H_{\textrm{total}} \mleft( \vec{y} \mright) - T_i \upDelta S_{\textrm{hyb}} \mleft( \vec{y} \mright) - \upDelta H_{\textrm{total}} \mleft( \vec{x} \mright) + T_i \upDelta S_{\textrm{hyb}} \mleft( \vec{x} \mright) \right) \notag \\
    & = \frac{\upDelta H_{\textrm{total}} \mleft( \vec{y} \mright) - \upDelta H_{\textrm{total}} \mleft( \vec{x} \mright)}{\kb T_j} - \frac{\upDelta H_{\textrm{total}} \mleft( \vec{y} \mright) - \upDelta H_{\textrm{total}} \mleft( \vec{x} \mright)}{\kb T_i} \notag \\
    & = \upDelta_{\textrm{c}} \mleft( \upDelta H_{\textrm{total}} \mright) \upDelta_{\textrm{r}} \beta,
\end{align}
\end{widetext}
where $\upDelta H_{\textrm{total}} \mleft( \vec{x} \mright) = \upDelta H_{\textrm{hyb}} \mleft( \vec{x} \mright) + \upDelta H_{\textrm{stack}} \mleft( \vec{x} \mright)$, with $\upDelta H_{\textrm{hyb}} \mleft( \vec{x} \mright)$, $\upDelta H_{\textrm{stack}} \mleft( \vec{x} \mright)$, and $\upDelta S_{\textrm{hyb}} \mleft( \vec{x} \mright)$ being the hybridization enthalpy, stacking energy, and hybridization entropy, respectively, for the selected model and system in configuration $\vec{x}$.
This allows us to use the values for these enthalpies and entropies that we update at each step without a full recalculation for different temperatures.
If instead only an additional bias term $U_\mathrm{bias}$ in the Hamiltonian is changing, then \cref{eq:remc-acc} simplifies further to
\begin{widetext}
\begin{equation}
    p_\textrm{acc} \mleft( \vec{y}; \mu_i, T_i, \mathcal{H}_i \land \vec{x}; \mu_j, T_j, \mathcal{H}_j \mid \vec{x}; \mu_i, T_i, \mathcal{H}_i \land \vec{y}; \mu_j, T_j, \mathcal{H}_j \mright) = \min \mleft[ 1,\ \textrm{e}^{\beta \upDelta_{\textrm{c}} \upDelta_{\textrm{r}} U_\mathrm{bias}} \mright]. \label{eq:remwus-acc}
\end{equation}
\end{widetext}

\ac{REMC} simulations were carried out for all four system.
In each case, three independent simulations were run, each with 16 replicas (and thus a range of 16 temperatures).
An iterative approach was used to refine the temperature selection such that there was an approximately equal spacing between the values of the averaged order parameters at the selected temperatures.
For systems S and D, the sharpness of the melting transition made selection of an appropriate temperature series challenging, as the melting temperature is not known a priori.
For accurate calculation of free energies, we instead ran \ac{US} simulations for these systems.

\subsection{Umbrella sampling}

We use a version of a \ac{MWUS} scheme~\cite{torrie1977_sup,mezei1987_sup,kastner2011_sup}.
In this scheme, a biasing potential must be chosen.
The optimal biasing potential for an order parameter $q$ is that which will give a uniform distribution,
\begin{equation}
    U_\mathrm{bias}(q) = \kb T \ln p (q),
\end{equation}
where $p (q)$ is the probability distribution of the order parameter in the given ensemble.
Of course, $p(q)$ is not known, and must be estimated in an iterative manner.
The order parameters $q$ here are all integer valued and fall within a relatively small range, so binning is not necessary to generate a histogram
After a set number of steps, the current simulation's histogram is used to estimate $U_\mathrm{bias}(q)$, which is then used as the bias weight in the next round.
To improve convergence during the early stages in which some bins may have very few samples and thus lead to poor estimates of $U_\mathrm{bias}(q)$, a maximum change in the bias weight is enforced.

To improve parallelization, rather than running a single simulation with the goal of achieving uniform sampling across the whole range of order parameters, multiple windows can be defined that cover only a subset of the range.
These windows have a further unvarying bias that prevents the simulation from sampling states outside the window; here a simple step function is used where the bias is zero inside the window range, and is determined by a linear function outside the window that slopes towards the region of zero bias.
In order to reconstruct a single free energy at the end of the simulation, the windows must overlap so that a single histogram may be constructed for the set of windows.
To improve efficiency, we additionally carry out replica-exchange steps between windows using \cref{eq:remwus-acc} as the acceptance criterion, which we refer to as \ac{REMWUS}.

For the two- and three-row systems, \ac{REMWUS} simulations were run at the melting temperature estimated via an iterative procedure.
The number of bound-domain pairs was used as the order parameter for the bias.
Nine windows were used; in the two-row system, these windows spanned three bound-domain pairs, while in the three-row system, they spanned four bound-domain pairs.
Starting configurations for each window were selected at random from the configurations produced by previous simulations (this was bootstapped by using configurations from Hamiltonian temperature \ac{REMC} simulations that themselves used a fully unbound system as a starting configuration).
Several iterations of the \ac{REMWUS} simulations were run to estimate the bias weights on the order parameter; the melting temperature was recalculated at the end, and this procedure itself was iterated with the updated melting temperature to improve the estimate, until a final production run was carried out.

\subsection{Calculation of free energies and expectation values}

We use the \ac{MBAR} method~\cite{shirts2008_sup} for simulation analysis.
The \ac{MBAR} method was developed to allow data from multiple simulations at different conditions to be combined, although in contrast to the \ac{WHAM}, it does not require the data to be binned to form histograms and provides an estimate of the uncertainty.
\ac{MBAR} may be used to reweight configurations to take advantage of data in conditions other than those of interest, as well as interpolate or extrapolate to conditions not actually simulated, if the change in weighting is not so great as to result in poor sampling of relevant states.

A series of independent samples is required as input.
We use the method of \citet{chodera2007_sup} to estimate the statistical inefficiency, which allows an uncorrelated subset of the samples generated by the simulations to be extracted.
Convergence of expectation values can be achieved with less data by discarding samples from the start of the simulation, when there are often highly atypical configurations.
We use an automated method for determining the equilibration, or burn-in, steps in the analysis of the \ac{REMC} simulations of systems S and D~\cite{chodera2016_sup}.
We used a freely available software package, pymbar, for the \ac{MBAR}, statistical inefficiency, and automated equilibration detection time calculations.
For the calculation of statistical inefficiency, we use a form of the reduced potential as the input series~\cite{shirts2008_sup},
\begin{equation}
    u_i \mleft( \vec{x} \mright) = \beta_i \left( U_i \mleft( \vec{x} \mright)+ \mu_i N \mleft( \vec{x} \mright) \right),
\end{equation}
where the indices refer to the particular state being considered.
The reduced potential is a suitable choice because it is a relatively general measure of relevant fluctuations in the system, and because it is also a required input of the \ac{MBAR} method.
In \cref{eq:chempot-ideal}, the chemical potential depends on the lattice constant; however, because the \ac{MBAR} method uses ratios of the exponent of the reduced potential, the lattice constant cancels out, and so does not need to be determined.

Three independent simulations were run for each condition presented in the paper, which were combined with the \ac{MBAR} analysis.
The error bars in all plots are the uncertainties given by the \ac{MBAR} analysis method.

To calculate the expectation values where we assume the binding domains act independently, we can simply use the \ac{NN} model.
Then, we can use the standard result for the chemical potential of ideal particles for the staples with amount concentration [A],
\begin{equation}
    \mu = \kb T \ln \mleft( \frac{[A]}{[\standardstate]} \mright),
\end{equation}
in \cref{eq:occupancy} to give the average occupancy
\begin{equation}
    \langle s \rangle = \frac{\mathrm{e}^{-\beta \dgsnn}}{\frac{[\standardstate]}{[\mathrm{A}]} + \mathrm{e}^{-\beta \dgsnn}}.
\end{equation}
The curves are calculated by multiplying $\langle s \rangle$ by the total number of staple types in the given system.
The entropy and enthalpy values that make up $\dgsnn$ are set to the averaged values used for a single bound-domain pair, the calculation of which is discussed below.
Because the contribution of $\dgsi$ is small and we are shifting the curves such that their melting temperatures become equal to those calculated from the simulations, for simplicity, we did not include it in the $\dgsnn$ values used here.

\subsection{Model parameters}

For all four systems studied here, we used a monovalent cation concentration of \SI{0.5}{\Molar}, a concentration of \SI{100}{\nano\Molar} for each staple type, and a default stacking energy of \SI{-1000}{\kB\kelvin}, which are the same parameters we focused on previously~\cite{cumberworth2018_sup}.
While the chemical potential also appears in the acceptance probability for staple exchange moves (see \citet{cumberworth2018_sup}), it can be shown that the lattice constant drops out, as the acceptance probability involves a single $\varepsilon_\mathrm{b}$, which also depends on the lattice constant (\cref{eq:bi-epsilon}).
Thus, we do not need to determine a value for the lattice constant.

All binding domains here are 16-\ac{nt}, which corresponds to 1.5 helical turns in a bound state, and so are modelled with the half-turn binding-domain potential.
For the sequence-specific simulations ran for systems S and D, the unified-\ac{NN} model~\cite{santalucia2004_sup} was used to calculate the hybridization free energies with \cref{eq:nn}; the sequences can be found in the replication package~\cite{replication-package_sup}.
The averaged hybridization free energies were calculated by using values for the \ac{NN} stacking enthalpies $\upDelta H_\mathrm{stack}^\standardstate$ and entropies $\upDelta S_\mathrm{stack}^\standardstate$ that were averaged over all ten possible nucleotide pairings, and multiplying by the number of \ac{NN} pairs per binding domain (15).
On average, one of the ends will have an AT base pair, so the corresponding \ac{NN} enthalpy $\upDelta H_\textrm{term}^\standardstate$ and entropy $\upDelta S_\textrm{term}^\standardstate$ penalty is added to the average.
The averaged entropy of hybridization was also corrected for a monovalent cation concentration of \SI{0.5}{\Molar} with \cref{eq:nn-salt}.
The averaged values for the misbound pairings were calculated by averaging over all possible misbound-domain pairs of the full tile system (which has a 56-binding-domain scaffold) that system D is a subset of~\cite{dannenberg2015_sup,dunn2015_sup}; the sequences for this system are also available in the replication package~\cite{replication-package_sup}.
Taking a simple mean over all misbound pairs may not give a good representation of misbinding, as a small number of much more favourable pairings may make a much larger contribution than the majority of pairings; however, we have chosen to start with this model for simplicity, and argue that this is not unreasonable based on our previous simulations, which find very little misbinding when using real sequences~\cite{cumberworth2018_sup}.
This averaging led to the binding enthalpy being set to \SI{-61000}{\kB\kelvin}, the binding entropy to \SI{-164}{\kB}, the misbinding enthalpy to \SI{-9100}{\kB\kelvin}, and the misbinding entropy to \SI{-24.2}{\kB}.
The initiation enthalpy $\upDelta H_\mathrm{init}^\standardstate$ and entropy $\upDelta S_\mathrm{init}^\standardstate$ are sequence independent, and so are unchanged.

    \putbib[si-tex/suppinfo,build/main_arxivNotes]
    \onecolumngrid
    \clearpage
    \pagebreak
    \section{Supplementary figures}

\begin{figure}[H]
    \centering
    \includegraphics{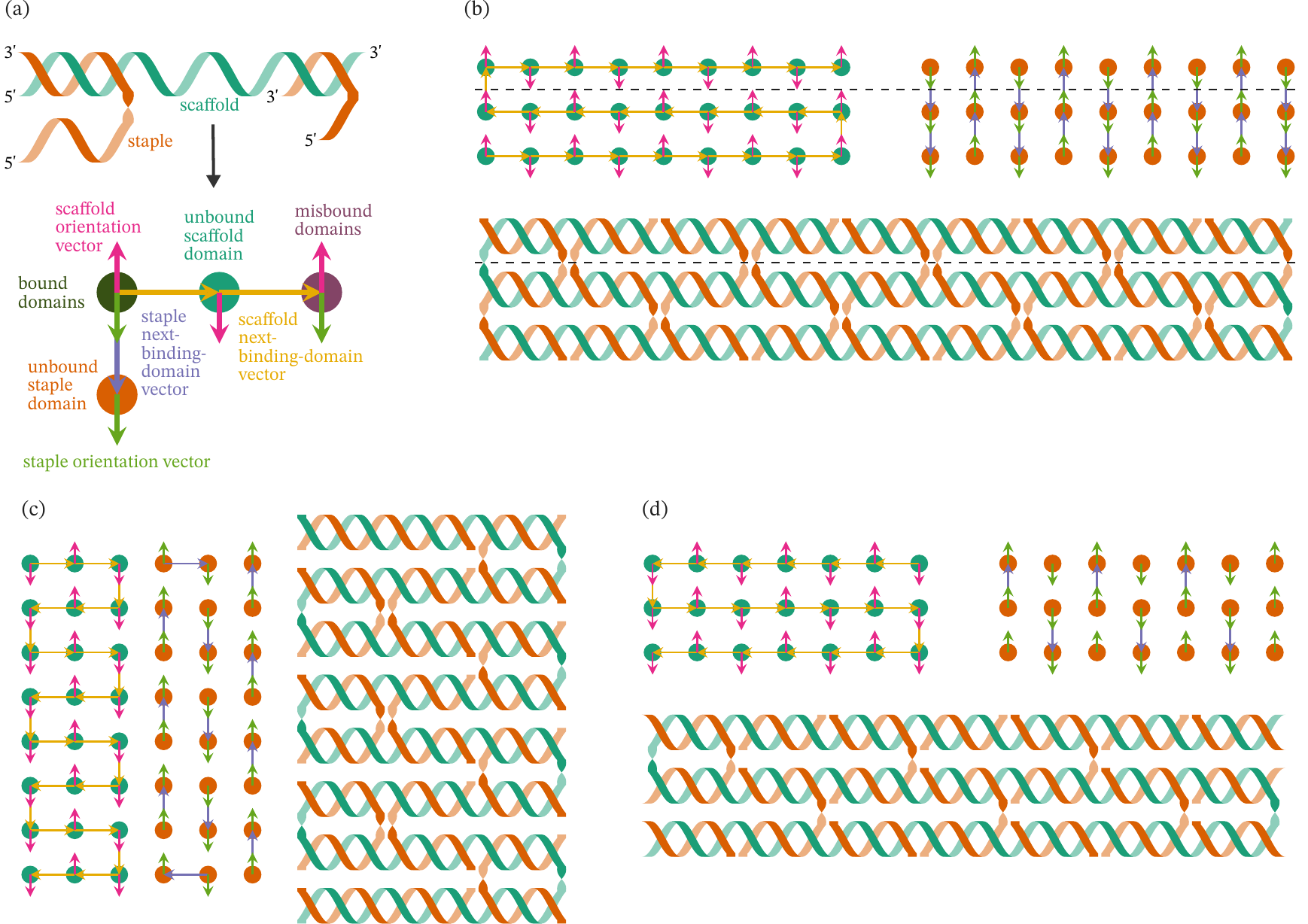}
    \caption{
        \label{fig:system-diagrams-sup}
        Representations of the systems simulated in this study.
        Each system is drawn with a cartoon helix representation and as the lattice model described in this work.
        (a) Legend showing a cartoon helix diagram above the lattice model representation.
        (b) The two- and three-row systems, with a dashed line showing the cut below which is the two-row system.
        (c) System S.
        (d) System D.
    }
\end{figure}

\begin{figure}[!p]
    \centering
    \includegraphics{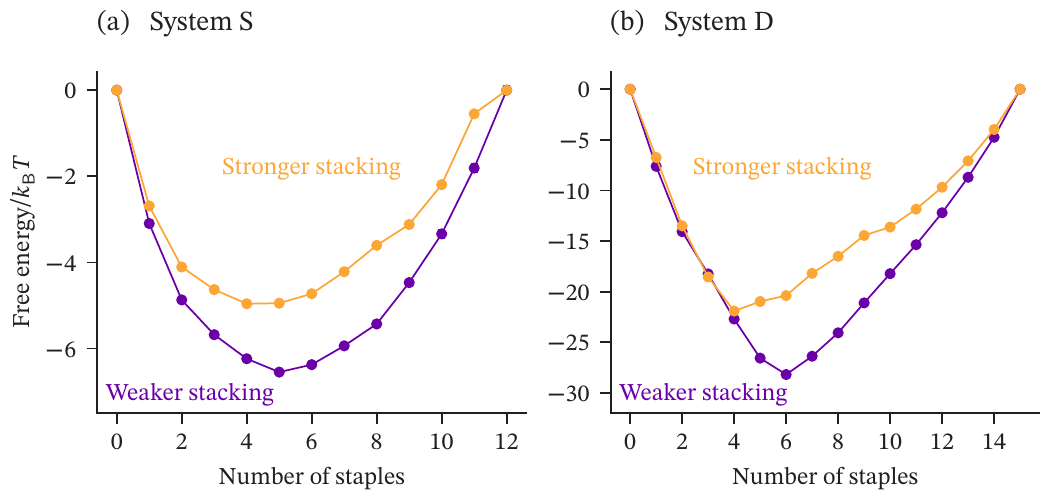}
    \caption{
        \label{fig:s-d-stacking}
        Free energies for the number of fully bound staples for system S (a) and system D (b) with both the default and double the coaxial stacking energy.
        Unlike in the case of the two-row system, no barrier emerges when the stacking energy is doubled in either system S or system D.
        This can be explained by the lower average number of stacking interactions per staple in these systems.
        System S has a relatively high number of edge staples, those which are at the ends of helices in the assembled state.
        Over half of the staples in system D are single-binding-domain staples, which in addition to having fewer stacking interactions, also bind at a relatively lower temperature than the two-binding-domain staples.
    }
\end{figure}

    \end{bibunit}
\end{document}